\documentclass[aps,prx,twocolumn,superscriptaddress,10pt,nofootinbib]{revtex4-2}
\usepackage{multirow,amsthm,amssymb,amsbsy,amsmath,epsfig,epstopdf,bm,float}
\DeclareMathOperator{\Tr}{Tr}
\usepackage{enumitem}
\usepackage{graphicx}
\usepackage{booktabs}
\usepackage{verbatim}
\usepackage{dcolumn}
\usepackage{color}
\usepackage[dvipsnames]{xcolor}
\usepackage{mathtools}
\usepackage[normalem]{ulem}
\usepackage{soul}
\usepackage{hyperref}
\usepackage{braket}
\usepackage[export]{adjustbox}
\usepackage{subfig}
\usepackage{float}
\theoremstyle{definition}
\newtheorem{definition}{Definition}

\theoremstyle{remark}

\theoremstyle{prop}
\newtheorem{prop}{Proposition}

\newcommand{\normgood}[1]{\left\lVert#1\right\rVert}

\DeclarePairedDelimiter\abs{\lvert}{\rvert}%

\newcommand{\mar}[1]{\textcolor{black}{#1}}

\usepackage{bm}

\newcommand\restr[2]{{
  \left.\kern-\nulldelimiterspace 
  #1 
  \vphantom{\big|} 
  \right|_{#2} 
  }}

\begin{document}

\title{
Quantum simulation of dissipative collective effects on noisy quantum computers
}

\author{Marco Cattaneo}
\email{marco.cattaneo@helsinki.fi}
\affiliation{QTF Centre of Excellence,  
Department of Physics, University of Helsinki, P.O. Box 43, FI-00014 Helsinki, Finland}
\affiliation{Instituto de F\'{i}sica Interdisciplinar y Sistemas Complejos (IFISC, UIB-CSIC), Campus Universitat de les Illes Balears E-07122, Palma de Mallorca, Spain}

\author{Matteo A. C. Rossi}
\affiliation{QTF Centre of Excellence, Department of Applied Physics, School of 
Science, Aalto University, FI-00076 Aalto, Finland}
\affiliation{Algorithmiq Ltd, Kanavakatu 3C 00160 Helsinki, Finland}

\author{Guillermo García-Pérez}
\affiliation{QTF Centre of Excellence,  
Department of Physics, University of Helsinki, P.O. Box 43, FI-00014 Helsinki, Finland}
\affiliation{Algorithmiq Ltd, Kanavakatu 3C 00160 Helsinki, Finland}
\affiliation{Complex Systems Research Group, Department of Mathematics and Statistics,
University of Turku, FI-20014 Turun Yliopisto, Finland}

\author{Roberta Zambrini}
\affiliation{Instituto de F\'{i}sica Interdisciplinar y Sistemas Complejos (IFISC, UIB-CSIC), Campus Universitat de les Illes Balears E-07122, Palma de Mallorca, Spain}

\author{Sabrina Maniscalco}
\affiliation{QTF Centre of Excellence,  
Department of Physics, University of Helsinki, P.O. Box 43, FI-00014 Helsinki, Finland}
\affiliation{QTF Centre of Excellence, Department of Applied Physics, School of 
Science, Aalto University, FI-00076 Aalto, Finland}
\affiliation{Algorithmiq Ltd, Kanavakatu 3C 00160 Helsinki, Finland}

\begin{abstract}

Dissipative collective effects are ubiquitous in quantum physics, and their relevance ranges from the study of entanglement in biological systems to noise mitigation in quantum computers. Here, we put forward the first fully quantum simulation of dissipative collective phenomena on a real quantum computer, based on the recently introduced multipartite collision model. 
First, we theoretically study the accuracy of this algorithm on near-term quantum computers with noisy gates, and we derive some rigorous error bounds that depend on the timestep of the collision model and on the gate errors. These bounds can be employed to estimate the necessary resources for the efficient quantum simulation of the collective dynamics. Then, we implement the algorithm on some IBM quantum computers to simulate superradiance and subradiance between a pair of qubits. Our experimental results successfully display the emergence of collective effects in the quantum simulation. In addition, we analyze the noise properties of the gates that we employ in the algorithm by means of full process tomography, with the aim of improving our understanding of the errors in the near-term devices that are currently accessible to worldwide researchers. We obtain the values of the average gate fidelity, unitarity, incoherence and diamond error, and we establish a connection between them and the accuracy of the experimentally simulated state. 
Moreover, we build a noise model based on the results of the process tomography for two-qubit gates and show that its performance is comparable with the noise model provided by IBM. 
Finally, we observe that the scaling of the error as a function of the number of gates is favorable, but at the same time reaching the threshold of the diamond errors for quantum fault tolerant computation may still be orders of magnitude away in the devices that we employ.

\end{abstract}
\maketitle

\section{Introduction}
Quantum simulation, i.e., the groundbreaking idea of simulating complex quantum {systems} on a {controllable} physical platform following the laws of quantum mechanics \cite{Feynman1982}, is probably the most promising application of quantum computers in the near future \cite{Altman2021}, owing to its exponential quantum advantage \cite{Lloyd1996} which would lead to crucial achievements in both fundamental and applied science \cite{Altman2021}. 
Quantum simulation of unitary many-body systems has been studied and experimentally implemented on several physical platforms \cite{Georgescu2014,Altman2021}, including superconducting quantum circuits \cite{Houck2012,Paraoanu2014,Kjaergaard2020,Blais2020}, trapped ions \cite{Blatt2012,Zhang2017,Monroe2021}, photonic systems \cite{Aspuru-Guzik2012} and cold atoms \cite{Bloch2012,Gross2017}. A relatively less studied problem is the quantum simulation of open quantum systems, whose evolution is not unitary due to the action of an external environment \cite{BreuerPetruccione,Rivas2012,weiss2012quantum}. Different protocols for open system quantum simulation have been introduced in the past twenty years \cite{Bacon2001,Lloyd2001,Koniorczyk2006,Wang2011,Kliesch2011,Muller2012a,Sweke2014,Sweke2015,Wei2016,Zanardi2016,Childs2017,Cleve2017,Patsch2020,Cattaneo2021,Schlimgen2021,Kamakari2021}, and careful studies of the resources they  require are available in the literature \cite{Kliesch2011,Sweke2014,Sweke2015,Childs2017,Cleve2017,Cattaneo2021}. On the experimental side, the main achievements include the simulation of a quantum map through quantum gates between trapped ions \cite{Barreiro2011,Schindler2013}, of different {single- and two-qubit} quantum channels on a near-term quantum computer \cite{Garcia-Perez2020}, of a single-qubit master equation via Trotterization in a superconducting quantum circuit \cite{Han2021}, of the Hubbard model with local dissipation on a near-term device \cite{Tornow2020}, of local fermionic reservoirs on non-interacting lattices \cite{DelRe2020,Rost2021}, and of local master equations via quantum imaginary time evolution on a near-term computer \cite{Kamakari2021}. In this work, we present the first fully quantum digital\footnote{That is, on a gate-based quantum computer.} simulation of dissipative collective effects on a real quantum computer, reproducing the dynamics driven by a \textit{global} master equation \cite{Hofer2017,Gonzalez2017,Cattaneo2019b} due to a common environment acting on the whole system. {The experiments have been run on near-term superconducting quantum computers available through the IBM cloud \cite{Qiskit}.}

Quantum dissipation is generally considered detrimental for
quantum technologies, because it inevitably induces decoherence on the system, hindering the realization of accurate quantum algorithms \cite{nielsenchuang}. Therefore, experimentalists usually try to reduce or counter dissipative processes in quantum computers. Here, our focus is different: we aim to engineer the most general coherent processes that occur in a dissipative global dynamics.
The importance of engineered collective dissipation is broad for both fundamental physics and quantum technologies. Simulating collective dissipation in a qubit platform {paves the way for} the experimental study of cutting-edge physical phenomena such as dissipative phase transitions \cite{Kessler2012}, quantum synchronization \cite{Giorgi2013,Walter2014,Xu2014} and dissipative time crystals \cite{Kessler2021}. Moreover, {such quantum simulations are an ideal experimental test-bed for} quantum thermodynamics, owing to the importance that global (i.e., collective) master equations have {in this field} \cite{Hofer2017,Gonzalez2017,Cattaneo2019b}. {Indeed, it has been shown that} local {master} equations may break the second law of thermodynamics \cite{Levy2014}, or modify the {related} energy contributions \cite{DeChiara2018a}. Engineered dissipation can also be used to build a different model of universal quantum computation \cite{Verstraete2009}, or to generate exotic entangled states by means of a common environment \cite{Braun2002,Lin2013,Kimchi-Schwartz2016}. In addition, collective phenomena are believed to play a major role in the dynamics of light-harvesting complexes \cite{Caruso2010}, therefore engineering global master equations may shed new light on the relation between dissipative quantum physics and biological systems. Last but not least, simulating collective dissipation would help us to detect and understand cross-talks in quantum computers \cite{Sarovar2020,VonLupke2020}, which may be one of the major sources of noise therein.

In this article, we investigate theoretically and experimentally one of the most well-known dissipative collective processes, {namely,} the emergence of superradiance \cite{Gross1982} and subradiance \cite{Crubellier1985} between qubits {emitting} simultaneously into a common environment. {For this purpose, we use} a quantum algorithm recently introduced {by some of us}, namely the \textit{multipartite collision model} (MCM) \cite{Cattaneo2021}. {The algorithm} reproduces the global emission by means of repeated interactions between the system qubits and a single ancillary qubit that mediates the collective decay. Collision (or repeated interactions) models are an important tool in the theory of open quantum systems \cite{Campbell2021a,Ciccarello2021,Cattaneo2022d}, with fundamental applications in quantum thermodynamics \cite{Barra2015,Strasberg2017,DeChiara2018a} and in the study of non-Markovianity \cite{Kretschmer2016a,Filippov2017}. Our work {is the first implementation of the MCM on a real quantum computer, and} one of the first experimental realizations of a collision model \cite{Cuevas2019,Garcia-Perez2020}.

{{Current quantum computers are} inevitably noisy \cite{Preskill2018}, limited by non-negligible gate errors and short coherence times, which {imposes strong constraints on the depth of} quantum circuits. {Hence, it is critical to analyze their limitations, identify possible improvements, and characterize their} errors \cite{Leymann2020}. In particular, near-term quantum computers are available on the cloud to a community of quantum physicists who aim to test their theories experimentally without having access to the hardware. Inevitably, this means that this community has  limited control over the characterization of the device and over the errors therein. It is our experience that the data about the noise features on the quantum computers available on the cloud are sometimes not sufficient to understand the errors we may face when running simulations on these devices. For instance, bad experimental results on the IBM quantum computers may not be justified by the gate errors provided by IBM \cite{Qiskit}. Starting from these considerations, we have decided to devote a considerable part of our work to} {studying, both theoretically and experimentally}, the errors arising {from} noisy gates. 

On the theoretical side, we  estimate a rigorous error bound for the precision of the quantum algorithm we have {used in order to} simulate dissipative collective effects in the presence of noise. On the experimental side, we perform process tomography \cite{Chuang1997,Poyatos1997,nielsenchuang} of all the gates employed in the algorithm, and state tomography of the system qubits at each step of the quantum simulation. {This allows us to better understand the type of noise affecting the quantum devices used in the experiments, to build a noise model based on the experimental gate process tomography,} and to propose possible countermeasures. 

A crucial aspect of our analysis {is} the choice of the figures of merit to estimate the noise properties.
Gate process tomography allows us to compute the \textit{average gate fidelity} \cite{Nielsen2002} of all the CNOTs we employ in the algorithm and their \textit{diamond distance} \cite{Kitaev1997,Watrous2018} from the ideal gate. These values capture the features of the error in each individual gate, but are also subject to the so-called \textit{SPAM errors}, i.e., errors arising from incorrect state preparation and measurement \cite{Merkel2013}. In contrast, the gate error provided by IBM is obtained through a protocol called \textit{randomized benchmarking} \cite{Knill2008,Magesan2011,Magesan2011a,Magesan2012a,Wallman2014,Wallman2018}, which estimates a sort of average error for single- and two-qubit gates on some selected qubits, but it is not subject to SPAM errors. 

We compare the experimental average gate infidelity with the IBM gate error, showing that in some cases the two values can be remarkably different. Two different noise models can be built based on these quantities, and we observe that their performance is similar. However, the noise model that makes use of the experimental process tomography may give better predictions in the presence of some highly noisy gates. 

The value of the diamond distance between ideal and experimental CNOT gates is also of interest to us for two main reasons. First, it directly relates to the theoretical error bound we estimate for the MCM with noisy gates. Secondly, it is the most used figure of merit to estimate the gate error thresholds for fault-tolerant quantum computation \cite{nielsenchuang}, {given that} the average gate fidelity is not reliable for this purpose \cite{Sanders2015}. Having a grasp on the magnitude of the diamond distance is of particular importance as the first experiments aimed at proving quantum advantage have been recently presented \cite{Arute2019,Wu2021}. Our results suggest that the errors in the near-term devices we have employed may still be orders of magnitude away from the strictest fault-tolerance thresholds.

The paper is structured as follows. In Sec.~\ref{sec:FiguresOfMerit} we introduce the most important theoretical tools to estimate the distance between quantum channels and the errors on a quantum computer. We also discuss the properties of the figures of merit used in our study and compare them. In Sec.~\ref{sec:multipartiteCM} we briefly recall the multipartite collision model and the dissipative quantum dynamics we aim to simulate. Sec.~\ref{sec:errorBoundTheo} is devoted to the discussion of the new theoretical error bounds for the quantum simulation of open systems with noisy gates. In Sec.~\ref{sec:IBM} we present in detail our experimental results and the noise analysis. Finally, in Sec.~\ref{sec:conclusions} we draw some concluding remarks and discuss the importance and significance of our results.

\section{Error estimation in quantum simulation}
\label{sec:FiguresOfMerit}
The goal of quantum simulation is to reproduce any quantum dynamics by means of a suitable composition of quantum channels that we can easily implement on a quantum computer. Therefore, {errors in quantum simulation algorithms arise from} the difference between {such ideal quantum channels and the physical ones implemented in practice}. In this section, {we} introduce some measures {to quantify the distance between} two quantum channels, {in order to provide a precise estimate of the} error of a quantum simulation algorithm. Specific emphasis  {is given to} the distance between the ideal quantum gate {of our theoretical} algorithm and {its} noisy version, {i.e., the} actual gate implemented on the quantum computer.

{We consider quantum channels $\mathcal{T}:\mathcal{B}(\mathcal{H})\rightarrow\mathcal{B}(\mathcal{H})$, with $\mathcal{B}(\mathcal{H})$ the space of bounded operators on the Hilbert space of the qubits. According to their usual definition, quantum channels are completely positive, linear, trace preserving maps \cite{nielsenchuang,BreuerPetruccione,Watrous2018}. }


\subsection{Distances between quantum channels and their properties}

{In this section we introduce the figures of merit for estimating the distance between two quantum channels. These are based either on the ``average'' or on the ``worst-case'' scenario \cite{Gilchrist2005}. Throughout the paper we will also employ the trace norm $\normgood{\cdot}_1$ and the infinity norm $\normgood{\cdot}_\infty$, which are recalled in Appendix~\ref{sec:distances}.}


\begin{definition}[Average gate fidelity]
If $\mathcal{U}_g$ is the (ideal) unitary superoperator associated with a quantum gate, and $\mathcal{T}$ is its noisy implementation, the average gate fidelity \cite{Nielsen2002} is defined as:
\begin{equation}
\label{eqn:averageGateFid}
\begin{split}
    \varphi(\mathcal{U}_g,\mathcal{T})&=\int d\mu(\psi)\mathcal{F}(\mathcal{U}_g[\ket{\psi}\bra{\psi}],\mathcal{T}[\ket{\psi}\bra{\psi}])\\
    &=\int d\mu(\psi) \bra{\psi}\!{\mathcal{U}_g^{-1}\mathcal{T}[\ket{\psi}\bra{\psi}]}\!\ket{\psi},
\end{split}
\end{equation}
where $d\mu(\psi)$ is the Haar measure over the pure states of the Hilbert space {and $\mathcal{F}$ is the fidelity between two quantum states introduced in Appendix~\ref{sec:distances}}. Correspondingly, we introduce the \textit{average gate infidelity} as $r(\mathcal{U}_g,\mathcal{T})=1-\varphi(\mathcal{U}_g,\mathcal{T})$. Note that if we average over the mixed states instead of pure states we obtain the same result, given the linearity in the definition of a density matrix \cite{Gilchrist2005} and the convexity of the infidelity \cite{Watrous2018}. 
\end{definition}

From an abstract perspective, obtaining the exact value of the average gate fidelity requires perfect knowledge on the channel $\mathcal{T}$, and this is obtained through standard quantum process tomography \cite{nielsenchuang}. The latter, however, has the drawback of being subject to errors in the state preparation and measurement in the circuits for the experimental process tomography \cite{Merkel2013}, and it is computationally feasible only for few-qubit gates. For this reason, a procedure called randomized benchmarking has been introduced \cite{Knill2008,Magesan2011,Magesan2011a,Magesan2012a,Wallman2014,Wallman2018}, which does not require full-state tomography, nor a precise control over the preparation and/or measurement errors. The core idea of the randomized benchmarking protocols is to apply a sequence of gates randomly drawn from the Clifford group and to compose it with its conjugate transpose on the selected qubits, so that, if the gates were ideal, the outcome channel would be equal to the identity. The ``survival probability'' of an initial state under this kind of evolution is then computed as a function of the number of gates in the sequence, and an average gate error is estimated by properly fitting the results (we refer the reader to extensive discussions in the literature for a more rigorous definition of what randomized benchmarking is actually measuring \cite{Magesan2012a,Proctor2017a,Wallman2016,Merkel2021}). Randomized benchmarking is gauge free and it is robust both to noise fluctuations on the ``twirling gates'' (the gates that form the Clifford group action) and to SPAM errors \cite{Wallman2016,Merkel2021}. However, it does not estimate the average fidelity of each individual gate \cite{Magesan2012a}. Individual gate fidelities may be estimated through \textit{interleaved randomized benchmarking} \cite{Magesan2012}, although some assumptions must be made for this protocol to be reliable \cite{Carignan-Dugas2019a}.

To the best of our knowledge, the single-qubit and two-qubit gate errors provided by IBM are obtained through standard randomized benchmarking \cite{Qiskit}. In the experimental part of this work, we will estimate the two-qubit average gate fidelity in Eq.~\eqref{eqn:averageGateFid} via process tomography and we will reduce the SPAM errors through readout error mitigation. A ``third way'' between process tomography and randomized benchmarking 
may be \textit{gate-set tomography} \cite{Merkel2013,Blume-Kohout2017a,DiMatteo2020,Nielsen2021}, which is a type of calibration-free tomography. However, it suffers from some so-called ``gauge issues'' \cite{Rudnicki2018,DiMatteo2020}, so we will not consider it here and leave it for future experimental studies.

\begin{definition}[Induced superoperator norm]
The $1\rightarrow 1$ superoperator norm of a quantum channel $\mathcal{T}$ is defined as:
\begin{equation}
\label{eqn:1to1supNorm}
\normgood{\mathcal{T}}_{1\rightarrow 1}=\max_{\normgood{\rho}_1=1}\normgood{\mathcal{T}[\rho]}_1.
\end{equation}
\end{definition}
The $1\rightarrow 1$ superoperator norm has the drawback of not
behaving well with respect to the tensor product. For this reason, a different norm is usually employed in quantum error correction algorithms:
\begin{definition}[Diamond norm]
The diamond norm of a quantum channel $\mathcal{T}$ is defined as \cite{Wilde2017,Watrous2018}:
\begin{equation}
\label{eqn:diamondNorm}
    \normgood{\mathcal{T}}_\Diamond=\normgood{\mathcal{T}\otimes\mar{\mathcal{I}}_A}_{1\rightarrow 1},
\end{equation}
where $\mar{\mathcal{I}}_A$ is the identity superoperator over a copy of the space $\mathcal{B}(\mathcal{H})$.
\end{definition}
Note that both the $1\rightarrow 1$ and the diamond norms are sub-multiplicative. Furthermore, the diamond norm satisfies the fundamental \textit{stability} property (see for instance Refs.~\cite{Kitaev2002,Gilchrist2005}), i.e., $\normgood{\mathcal{T}\otimes\mar{\mathcal{I}}_B}_\Diamond=\normgood{\mathcal{T}}_\Diamond$, where $\mar{\mathcal{I}}_B$ is the identity superoperator over a generic space. Specifically, it can be shown that the diamond norm defined with the identity superoperator over a copy of the space $\mathcal{B}(\mathcal{H})$ is the maximal one (it is lower for lower dimensions, and it has the same values for higher dimensions). The diamond norm can be computed either through convex optimization procedures \cite{Ben-Aroya2010} or by a semidefinite program \cite{Watrous2012,Watrous2011}. 

\begin{definition}[Diamond distance] The diamond distance between an ideal gate with channel $\mathcal{U}_g$ and its noisy realization $\mathcal{T}$ is defined as: 
\begin{equation}
\label{eqn:diamondDistance}
    d_\Diamond(\mathcal{U}_g,\mathcal{T})=\frac{1}{2}\normgood{\mathcal{U}_g-\mathcal{T}}_\Diamond.
\end{equation}
\end{definition}
The sub-multiplicativity of the diamond norm implies that:
\begin{equation}
\label{eqn:submultiplicativity}
    \normgood{\mathcal{U}_1\mathcal{U}_2-\mathcal{T}_1\mathcal{T}_2}_\Diamond\leq\normgood{\mathcal{U}_1-\mathcal{T}_1}_\Diamond+\normgood{\mathcal{U}_2-\mathcal{T}_2}_\Diamond,
\end{equation}
and analogously for the $1\rightarrow 1$ superoperator norm.
This is the \textit{sub-additivity} property, which is crucial to guarantee that the error in the gate composition scales at most linearly as a function of the length of the gate sequence.

Note that both the $1\rightarrow 1$ and the diamond norm do not have a closed form as a function of the Choi matrix of the quantum channel we are interested in, but must be computed through numerical maximization. Moreover, to obtain their values we must typically rely on quantum process tomography, which is sensitive to SPAM errors (while, for instance, the average gate fidelity may be estimated through a SPAM-free randomized benchmarking protocol).

In addition, we introduce a figure of merit to estimate the \textit{coherence of noise} \cite{Wallman2015a}. That is, {how close a given quantum channel $\mathcal{T}$ is to a unitary one}. This will be crucial to understand whether the error of a noisy quantum gate will be mostly due to the presence of decoherence or to the fact that we are performing a gate which is different from the target gate (say, we perform \mar{a qubit rotation driven by $\sigma_x^{0.99}$ instead of the desired $\sigma_x$ or, more drastically, a rotation driven by $\sigma_z$ instead of $\sigma_x$}).
\begin{definition}[Unitarity]
The unitarity of a quantum channel $\mathcal{T}$ is defined as the average purity of output states, with the identity components subtracted:
\begin{equation}
\label{eqn:unitarity}
    u(\mathcal{T})=\frac{d_S}{d_S-1}\int {d\mu(\psi)\Tr\left[\mathcal{T}\left[\ket{\psi}\bra{\psi}-\frac{\mathbb{I}_S}{d_S}\right]^2\right]},\\
\end{equation}
where $d_S$ is the dimension of the system Hilbert space.
\end{definition}

Equivalently, the unitarity can be defined as \cite{Wallman2015a,Carignan-Dugas2019a}
\begin{equation}
    \label{eqn:unitarityBis}
    u(\mathcal{T})=\frac{1}{d_S^2-1}\Tr[\mathcal{T}_u^\dagger\mathcal{T}_u],
\end{equation}
where $\mathcal{T}_u$ is the \textit{unital block} \cite{Carignan-Dugas2019a}, defined by the following Liouvillian representation of the trace-preserving channel:
\begin{equation}
\label{eqn:rapUn}
    \mathcal{T}=\begin{pmatrix}
    1 & 0 \\
    \mathcal{T}_n & \mathcal{T}_u
    \end{pmatrix}. 
\end{equation}
The orthonormal basis in which we are representing the matrix is $\{B_0=\mathbb{I}_S/\sqrt{d_S},B_1,B_2,\ldots,B_{d_S^2-1}\}$ with $\Tr[B_j^\dagger B_k]=\delta_{jk}$, thus $\Tr[B_k]=0$ for $k\geq 1$. For instance, for qubit systems we will choose the basis given by the tensor products of Pauli matrices, including the identity. $\mathcal{T}_n$ is the \textit{non-unital vector} \cite{Carignan-Dugas2019a}.

The unitarity of a general quantum channel satisfies $u(\mathcal{T})\leq 1$, being one only if $\mathcal{T}$ is unitary. Moreover, $u(\mathcal{V}\mathcal{T}\mathcal{U})=u(\mathcal{T})$ for all $\mathcal{U},\mathcal{V}$ unitary. The unitarity satisfies the following lower bound in terms of the average gate infidelity: $u(\mathcal{T})\geq [1-d_S r(\mathcal{U}_g,\mathcal{T})/(d_S-1)]^2$ for any $\mathcal{U}_g$ (clearly, the closer $\mathcal{U}_g$ is to $\mathcal{T}$ the tighter the bound \cite{Wallman2015a}).

Finally, we will make use of the \textit{incoherence}  \cite{Feng2016,Yang2019}, which is a quantity related to the unitarity as follows:
\begin{definition}[Incoherence]
The incoherence of a channel $\mathcal{T}$ is defined as:
\begin{equation}
    \label{eqn:incoherence}
    \omega(\mathcal{T})=\frac{d_S-1}{d_S}(1-\sqrt{u(\mathcal{T})}).
\end{equation}
\end{definition}
Using the lower bound for the unitarity, we readily obtain $0\leq\omega(\mathcal{T})\leq r(\mathcal{U}_g,\mathcal{T})$ for any $\mathcal{U}_g$. The value of the incoherence can be thought of as the minimum infidelity that may be achieved with perfect unitary control over the system \cite{Yang2019}, i.e., the ``contribution'' to the infidelity due to purely dissipative errors. A different measure of the ``non-unitarity'' of a quantum channel has also been introduced in the literature \cite{Das2018}, but for simplicity it won't be considered in this study.

\subsection{Comparing different figures of merit}
\label{sec:comparingFig}
It is a well-known fact that the fidelity is a deceptive measure to compare quantum states. For instance, the fidelity between states with very different properties, e.g., one entangled and the other separable, can be larger than 0.95 \cite{Bina2014}. This is a consequence of the infidelity not being a proper mathematical distance. If we compare it to a well-defined distance such as the trace norm, a common inequality reads \cite{Watrous2018}:
\begin{equation}
\label{eqn:fidelityBound}
    1-\sqrt{\mathcal{F}(\rho,\sigma)}\leq \frac{1}{2}\normgood{\rho-\sigma}_1\leq \sqrt{1-\mathcal{F}(\rho,\sigma)},
\end{equation}
where, crucially, the trace norm is bounded by the square root of the infidelity. This means that states with fidelity 0.99 may have a trace distance equal to 0.1.

We can expect similar considerations also when the fidelity is employed to estimate the distance between quantum channels, as in Eq.~\eqref{eqn:averageGateFid}. Indeed, the most important and rigorous theorems guaranteeing fault-tolerant quantum computation do not rely on the fidelity, and {instead} make use of the diamond distance to bound the maximal error which can be cast away by means of quantum error correction protocols \cite{Kitaev1997,Aharonov2008,Sanders2015,Kueng2016}. The diamond norm is preferred to the $1\rightarrow 1$ superoperator norm because of its stability property, discussed in the previous section. Moreover, addressing the worst case scenario (i.e., performing a maximization as in Eq.~\eqref{eqn:diamondNorm}) is necessary to estimate a proper fault-tolerant error threshold, while one cannot rely on the average figure of merit in Eq.~\eqref{eqn:averageGateFid}. 

As a possible alternative to average gate fidelity, the \textit{entanglement fidelity} \cite{Schumacher1996,Nielsen1996,Reimpell2005,Horodecki2009} of a quantum channel has been proposed to estimate the noise properties of a quantum channel.
\begin{definition}[Entanglement fidelity] 
The entanglement fidelity of a quantum channel $\mathcal{T}$ is defined as:
\begin{equation}
    \label{eqn:entanglementFidelity}
    F_e(\mathcal{T})=\bra{\phi}(\mathcal{T}\otimes\mar{\mathcal{I}}_A)[\phi]\ket{\phi},
\end{equation}
where $\mar{\mathcal{I}}_A$ is the identity \mar{superoperator} acting on a copy of the Hilbert space of the system, while $\phi$ represents a maximally entangled state in the extended Hilbert space $\mathcal{H}\otimes\mathcal{H_A}$.
\end{definition}
Entanglement fidelity captures how well entanglement between the quantum system of interest and other systems is preserved under the local application of the channel $\mathcal{T}$. It has been shown that the entanglement fidelity can be directly connected to the average gate fidelity as \cite{Horodecki1999,Nielsen2002}:
\begin{equation}
    \label{eqn:entFidAvGateFid}
\varphi(\mathcal{U}_g,\mathcal{T})=\frac{d_S F_e(\mathcal{U}_g^\dagger \mathcal{T})+1}{d_S+1},
\end{equation}
where $d_S$ is the dimension of the system Hilbert space.
\mar{However, the entanglement fidelity is not
a measure of the worst case error either.}

{An} upper bound for the diamond distance as a function of the system dimension $d_S$ and the average gate infidelity $r(\mathcal{U}_g,\mathcal{T})$ {is given by} \cite{Wallman2014}:
\begin{equation}
\label{eqn:upperBoundDiamond}
d_\Diamond(\mathcal{U}_g,\mathcal{T})\leq d_S \sqrt{(1+d_S^{-1})(r(\mathcal{U}_g,\mathcal{T}))}.
\end{equation}
Once again, we see that the square root of the average gate infidelity appears in the bound. Moreover, the dependence on the system dimension makes its scaling even worse than in Eq.~\eqref{eqn:fidelityBound}. The tightness of this bound has not been proven, but some results {indicate that} the bound is asymptotically tight as a function of $\varphi$ and $d_S$ \cite{Sanders2015}. They also show how, in general, the average gate fidelity is not reliable for assessing fault-tolerant quantum computation: a value of $\varphi(\mathcal{U}_g,\mathcal{T})=99\%$ \mar{for a two-qubit gate} can still lead to an error around $\epsilon=0.13$, which is {very} far from the fault-tolerant thresholds against general noise (typically $\epsilon\approx 10^{-3}$--$10^{-4}$ \cite{Blume-Kohout2017a,Aliferis2009,Jones2018,Puri2020,Nielsen2021}).

Therefore, the diamond norm is the only reliable figure of merit for estimating the distance between quantum gates with the aim of assessing fault-tolerance of the quantum computation, while the value of the average gate fidelity, which is very useful to estimate the average error we will run into during the implementation of a quantum gate \cite{Nielsen2002,Gilchrist2005}, must be taken with a grain of salt when speaking about fault-tolerance (more discussions can be found in Refs.~\cite{Wallman2014,Sanders2015,Wallman2015a,Wallman2015,Kueng2016}). 

A tighter error bound for the diamond distance can be employed if we know the unitarity of the quantum channel \cite{Wallman2015} (a similar bound can be found in Ref.~\cite{Kueng2016}):
\begin{equation}
    \label{eqn:upperBoundDiamondUnitarity}
    d_\Diamond(\mathcal{U}_g,\mathcal{T})\leq \sqrt{\frac{d_S^3 C^2}{4}+\frac{(d_S+1)^2 (r(\mathcal{U}_g,\mathcal{T}))^2}{2}},
\end{equation}
with $C^2=\frac{d_S^2-1}{d_S^2}(u(\mathcal{T})-2p(\mathcal{U}_g,\mathcal{T})+1)$ and $p(\mathcal{U}_g,\mathcal{T})=1-(d_S r(\mathcal{U}_g,\mathcal{T}))/(d_S-1)$.

Finally, the unitarity can be used to estimate how the average gate fidelity of composed channels behaves as a function of the average gate fidelity of each component \cite{Carignan-Dugas2019a}, {or equivalently, how the average gate fidelity scales as a function of the number of gates in the algorithm}. 
Without getting into the details discussed in Ref.~\cite{Carignan-Dugas2019a}, if we consider, for simplicity, a circuit made of $m$ identical gates with average gate fidelity $\varphi$, then the average gate infidelity of the whole circuit will be upper bounded by $m^2 (1-\varphi)$, plus some lower-order terms that scale as $m^4(1-\varphi)^2$ \footnote{Note that a universal bound for any $m$ and $\varphi$ cannot be found, because the average gate infidelity cannot be higher than 1}. This bound is tight for all unitary channels with unitarity $u=1$. If, on the contrary, one considers highly incoherent channels, such as the depolarizing one, the bound can be improved as $m(1-\varphi)$, plus terms of the order of $O(m^2(1-\varphi)^2)$. {To summarize,} lower unitarity implies a better scaling of the infidelity as a function of the channel length.

\section{Multipartite collision model for dissipative collective effects}
\label{sec:multipartiteCM}
\subsection{Introduction to the algorithm}
\label{sec:algorithm}
The multipartite collision model (MCM) has been recently introduced \cite{Cattaneo2021} as a repeated interaction model \cite{Campbell2021a,Ciccarello2021} able to reproduce any Markovian evolution of a multipartite open quantum system (that is, an open system made of multiple subsystems) by elementary collisions between each subsystem and some environment particles, termed as ``ancillas''. 
{Collision models such as the MCM are naturally suited to implement digital quantum simulations of open quantum dynamics.}
{Specifically, we consider} a quantum system living in a Hilbert space $\mathcal{H}$ composed of $M$ subsystems, which, {without loss of generality}, we consider identical and of dimension $d$:
\begin{equation}
\label{eqn:HilbertSpace}
\mathcal{H}=\bigotimes_{j=1}^M \mathcal{H}_j, \quad\dim(\mathcal{H}_j)=d \quad\forall j.
\end{equation}
We are interested in the most general Markovian quantum evolution of the state of the system at time $t$, $\rho_S(t)$. Specifically, starting from a generic $\rho_S(0)$, we aim to simulate the dynamics
\begin{equation}
\label{eqn:evolution}
    \rho_S(t)=\exp\mathcal{L}t[\rho_S(0)]
\end{equation}
on a quantum computer, with $\mathcal{L}$ is the so-called Liouvillian superoperator, i.e., the generator of the quantum dynamical semigroup driving the {dynamics}. The most general structure of the Liouvillian is expressed by the celebrated Gorini-Kossakowski-Sudarshan-Lindblad (GKLS) master equation \cite{BreuerPetruccione}:
\begin{equation}
\label{eqn:gkls}
    \mathcal{L}[\rho_S(t)]=\frac{d}{dt}\rho_S(t)=-i[H_S,\rho_S(t)]+\mathcal{D}[\rho_S(t)],
\end{equation}
where we are working with units such that $\hbar=1$. $H_S$ is a generic Hermitian operator of the system, to which we refer as the ``effective Hamiltonian'', while
\begin{equation}
\label{eqn:dissipator}
    \mathcal{D}[\rho_S]=\sum_{k=1}^{J} \Gamma_k\left( L_k \rho_S L_k^\dagger-\frac{1}{2}\{L_k^\dagger L_k,\rho_S\}\right)
\end{equation}
is the dissipator, describing the non-unitary system dynamics. $\Gamma_k\geq 0$ are non-negative decay rates, while $\{L_k\}_{k=1}^{J}$ are the Lindblad operators, whose number $J$ is bounded as $J\leq d^{2M}-1$.

The multipartite collision model provides a way to simulate Eq.~\eqref{eqn:evolution} on a quantum computer by expressing the action of the dissipator {of} Eq.~\eqref{eqn:dissipator} in terms of elementary quantum gates between each subsystem, {i.e., a subset of qubits}, and a collection of ancillary qubits, {representing} the environment ancillas. Dissipative collective effects, {due to} terms in Eq.~\eqref{eqn:dissipator} that couple two or more subsystems, {are implemented via} a suitable sequence of quantum gates between a single environment ancilla and two or more system qubits. More precisely, collective terms are treated through a second-order Suzuki-Trotter decomposition \cite{Hatano2005}. A detailed presentation of {the MCM algorithm} {is summarized} 
in Appendix~\ref{sec:steps}.

The fundamental result introduced in~\cite{Cattaneo2021} states that the MCM simulates the exact open system dynamics driven by $\mathcal{L}$ (Eq.~\eqref{eqn:evolution}) in the limit of small timestep $\Delta t$:
\begin{equation}
\label{eqn:simulation_result}
    \lim_{\Delta t\rightarrow 0^+}(\phi_{\Delta t})^n=\exp\mathcal{L}t,\qquad n=t/\Delta t,
\end{equation}
where $\phi_{\Delta t}$ is the quantum channel reproducing a single step of the MCM (see Appendix~\ref{sec:steps} for details){, which here has been applied $n$ times (i.e., $n$ repeated collisions until time $t=n\Delta t$)}.

On a real quantum computer, obviously one can not employ an infinitesimal timestep $\Delta t\rightarrow 0^+$ to implement the gates in Eq.~\eqref{eqn:suzuki-Trotter}, and has to choose a small but finite $\Delta t$. This implies that the multipartite collision model is able to simulate the dynamical semigroup as in Eq.~\eqref{eqn:simulation_result} up to an error that depends on $\Delta t$. Let us term this global error as $\epsilon_g$, following the original paper \cite{Cattaneo2021}:
\begin{equation}
    \label{eqn:global11}
    \epsilon_g=\normgood{(\phi_{\Delta t})^n-\exp\mathcal{L}t}_{1\rightarrow 1}.
\end{equation}
In the above equation we are using the $1\rightarrow 1$ superoperator norm despite its stability issues discussed in Sec.~\ref{sec:FiguresOfMerit} because this figure of merit was studied in Ref.~\cite{Cattaneo2021}. This definition will be extended through the use of the diamond norm in Sec.~\ref{sec:errorBoundTheo} and Appendix~\ref{sec:proofs}.

The global error can be trivially bounded by the sum of the errors in a single timestep: $\epsilon_g\leq n \epsilon_s$, where
\begin{equation}
    \label{eqn:errorSingle11}
    \epsilon_s=\normgood{\phi_{\Delta t}-\exp\mathcal{L}\Delta t}_{1\rightarrow 1}.
\end{equation}
Ref.~\cite{Cattaneo2021} estimates a very general bound on $\epsilon_s$, showing that its scaling is optimal for collision models. {This allows for the efficient implementation of the quantum simulation algorithm}. The complete expression of this bound is quite cumbersome and can be found in the Supplemental Material of Ref.~\cite{Cattaneo2021}. Here, we just write it as:
\begin{equation}
\label{eqn:bound}
\begin{split}
    \epsilon_s&\leq\mathcal{B}_{1\rightarrow 1}\\&=  2e(M\Lambda(1+JR\Lambda)\Delta t)^2\\&+\text{pol}_1(\Lambda,M,J)\Delta t^2+\text{pol}_2(\Lambda,M,J)\Delta t^3,
\end{split}
\end{equation}
where $\Lambda=\max_{k,m}\{\normgood{H_S}_\infty,\normgood{(\lambda_k F_m^{(k)}\sigma_k^++h.c.)}_\infty\}$ (the latter term being the Hamiltonian driving a single collision, as explained in Appendix~\ref{sec:steps}), and $\text{pol}_1$ and $\text{pol}_2$ are polynomial functions of $\Lambda$, the number of subsystems $M$, and the maximum number of jump operators $J$. The scaling of the global error is $\epsilon_g\sim O(n\Delta t^2)=O(t^2/n)$. Finally, note that $J$ scales exponentially with the number of subsystems. However, under the very common assumption of $k$-locality \cite{Lloyd1996,Kliesch2011}, i.e., the jump operators and the effective Hamiltonian can be decomposed as sums of $k$-local terms that act non-trivially on $k$ subsystems only, the bound in Eq.~\eqref{eqn:bound} scales polynomially as a function of $M$ \cite{Kliesch2011,Cattaneo2021}. This guarantees that the multipartite collision model is efficiently simulable on a quantum computer.

\subsection{Simulation of super- and sub-radiance}
\label{sec:subSup}
We will now show how the multipartite collision model can be applied to the simulation of topical collective effects, such as super- and sub-radiance. The simplest model where these phenomena arise consists of two atoms that emit coherently into a common environment. Specifically, superradiance emerges when the atoms lose their energy through a quick emission, the intensity of which is enhanced with respect to the local incoherent decay that each atom would experience in the absence of the other \cite{Gross1982}. In contrast, subradiance, which is a complementary effect due to the same cause (namely, the action of a common bath), can be identified as the presence of a slowly decaying, metastable mode of the atomic emission, which persists for a time way longer than the usual relaxation time  \cite{Crubellier1985} ($T_1$ in the language of cavity or circuit quantum electrodynamics \cite{Blais2020}). 

\begin{figure*}
    \centering
\subfloat[]{%
  \includegraphics[scale=0.4]{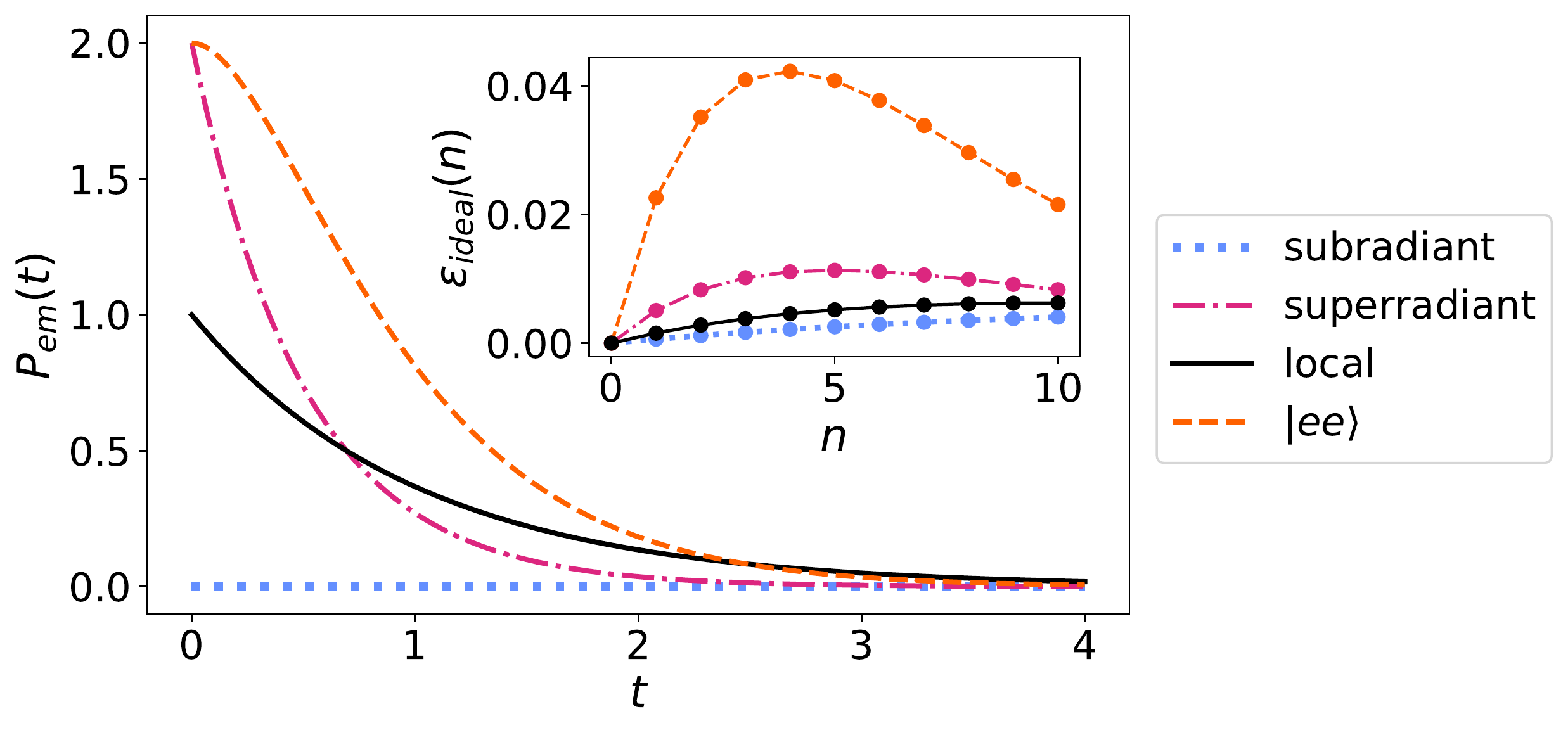}%
}
\subfloat[]{%
  \includegraphics[scale=0.4]{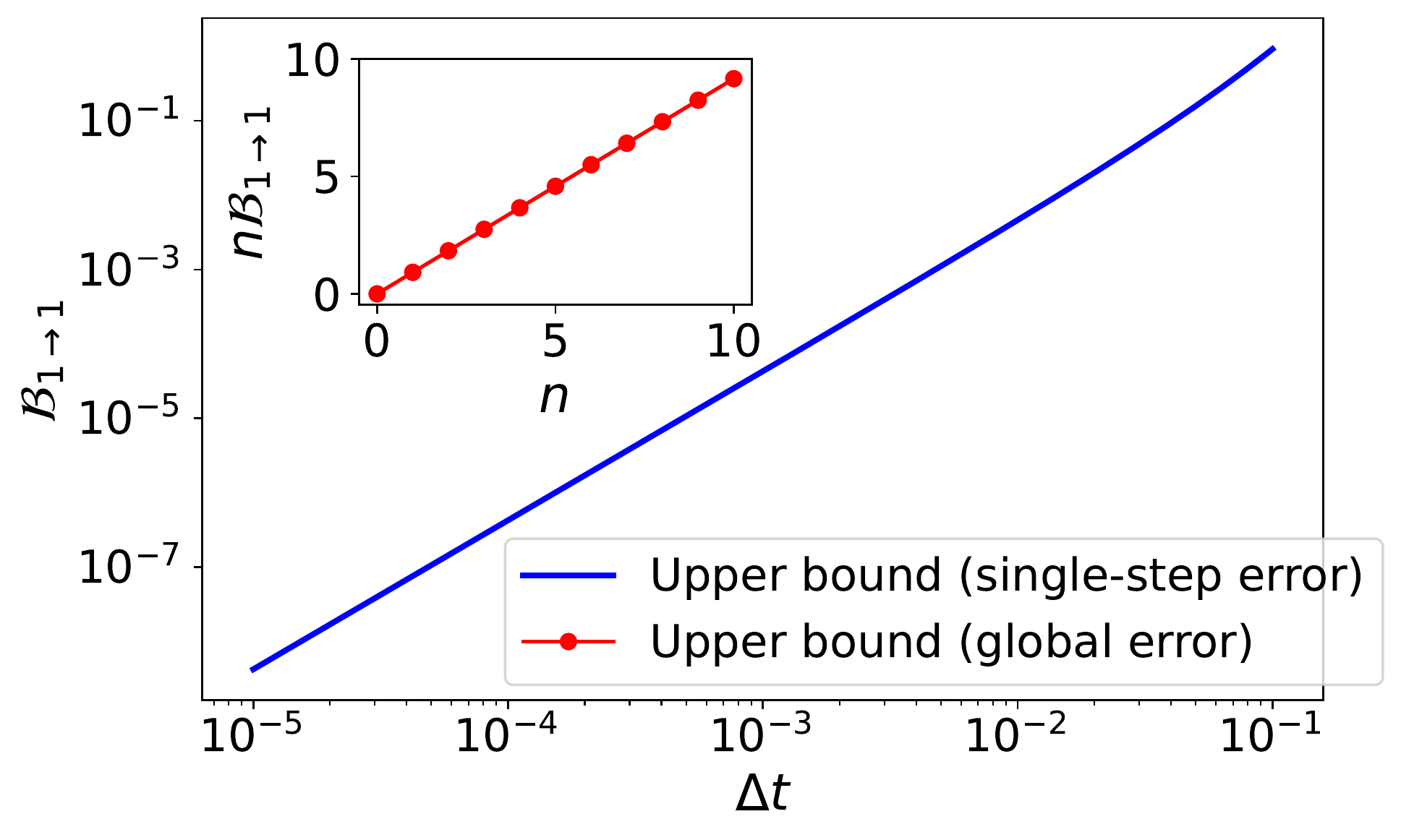}%
}
    \caption{(a): Emission power as a function of time, when the initial state is the subradiant state (dotted light blue line), superradiant state (dash-dotted magenta line), $\ket{ee}$ (dashed orange line), and when the dynamics is local starting from the subradiant state (solid black line). Inset in (a): Trace distance between real and simulated state $\epsilon_{\mathrm{ideal}}(n)$ as a function of the number of timesteps $n$, with the same initial states as in the main figure. (b): Upper bound $\mathcal{B}_{1\rightarrow 1}$ for the single-step error $\epsilon_s$ as a function of the timestep $\Delta t$. Inset in (b): Upper bound for the global error $\epsilon_g$ with $\Delta t=0.1$, equal to $n$ times $\mathcal{B}_{1\rightarrow 1}$, where $n$ is the number of timesteps.}
    \label{fig:ideal}
\end{figure*}

In this work we address super- and sub-radiance between two two-level atoms, i.e., two qubits of a quantum computer. Despite its simplicity, this model displays a rich landscape of collective phenomena \cite{Bellomo2017} that may also bring new insights on the presence of entangling noise in the experimental platform \cite{Cattaneo2021b}. Remarkably, we will address a scenario where subradiance emerges through the presence of a decoherence-free subspace, that is, a subspace of the system Hilbert space where the dynamics is unitary and there is no dissipation. The relevance of decoherence-free subspaces for quantum computation is crucial, given that noiseless computation may be possible by restricting the Hilbert space over which we run the quantum algorithms to the decoherence-free subspace \cite{Lidar2003}. More specifically, we aim to simulate the following generator of the open system dynamics:
\begin{equation}
\label{eqn:LiouvillianSup}
    \begin{split}
    \mathcal{L}[\rho_S]={}& \sum_{j,k=1,2}\gamma\left(\sigma_j^-\rho_S\sigma_k^+-\frac{1}{2}\{\sigma_k^+\sigma_j^-,\rho_S\}\right)\\
    ={}& \gamma\left( L_{col}\rho_SL_{col}^\dagger-\frac{1}{2}\{L_{col}^\dagger L_{col},\rho_S\}\right),
    \end{split}
\end{equation}
with $L_{col}=(\sigma_1^-+\sigma_2^-)$, and the latter are the lowering operators of respectively qubit $1$ and $2$. In Eq.~\eqref{eqn:LiouvillianSup} $\gamma$ is a constant decay rate that defines the magnitude of the dissipation {and, specifically, it is the spontaneous emission coefficient of each atom in the vacuum}. 

The master equation~\eqref{eqn:LiouvillianSup} can be derived from a microscopic model with two identical two-level atoms dissipatively coupled in a symmetric way to the same common bath at zero temperature \cite{Cattaneo2019b}. {In a reference frame rotating with the frequency of the qubits, the free Hamiltonian of the atoms can then be neglected,} as in Eq.~\eqref{eqn:LiouvillianSup}. 
The analytical solution of the dynamics driven by Eq.\eqref{eqn:LiouvillianSup} can be found in Appendix~\ref{sec:anSol}.

Superradiance is observed when the open evolution of the two qubits starts from the ``superradiant state'' $\ket{\psi_{\mathrm{sup}}}=\frac{\ket{ge}+\ket{eg}}{\sqrt{2}}$, where $\ket{g}$ and $\ket{e}$ are respectively the ground and the excited state of each qubit. Indeed, introducing $\rho_{\mathrm{sup}}=\ket{\psi_{\mathrm{sup}}}\bra{\psi_{\mathrm{sup}}}$ and the emission power \cite{Gross1982}
\begin{equation}
\label{eqn:emiPower}
    P_{em}[\rho_S(0)](t)=-\Tr[H \mathcal{L}[\exp\mathcal{L}t[\rho_S(0)]]],
\end{equation}
where $H=\frac{1}{2}(\sigma_1^z+\sigma_2^z)$ is the system's energy, we easily find (see Appendix~\ref{sec:anSol} for details):
\begin{equation}
\label{eqn:intensity}
    P_{em}[\rho_{\mathrm{sup}}](t)=2 \gamma e^{-2\gamma t}.
\end{equation}
We observe a clear enhancement of the atomic decay, driven by a decay rate equal to $2\gamma$ instead of the standard spontaneous emission rate $\gamma$.

In contrast, subradiance emerges when the system dynamics starts from the ``subradiant state'' $\ket{\psi_{\mathrm{sub}}}=\frac{\ket{eg}-\ket{ge}}{\sqrt{2}}$. Defining $\rho_{\mathrm{sub}}=\ket{\psi_{\mathrm{sub}}}\bra{\psi_{\mathrm{sub}}}$, we observe
\begin{equation}
    \mathcal{L}[\rho_{\mathrm{sub}}]=0,
\end{equation}
with zero emitted power.
In other words, the subradiant state is a steady state of the dynamics that does not ``feel'' the presence of dissipation acting on the qubits. 

Let us now expose how the multipartite collision model can simulate the dynamics in Eq.~\eqref{eqn:LiouvillianSup}. Since there is a single Lindblad operator ($J=1$), we only need a single ancillary qubit initialized in the ground state for each timestep $\Delta t$. Let us term it as the qubit $E$. Then, the two-qubit quantum gates we will employ in the algorithm are:
\begin{equation}
\label{eqn:gateSupSub}
    \begin{split}
        &U^{(1)}(\Delta t)=\exp\left[-i\sqrt{\Delta t}(\sqrt{\gamma}\sigma_1^-\sigma_E^++h.c.)\right],\\
        &U^{(2)}(\Delta t/2)=\exp\left[-i\frac{\sqrt{\Delta t}}{2}(\sqrt{\gamma}\sigma_2^-\sigma_E^++h.c.)\right].\\
    \end{split}
\end{equation}
These gates will be composed as in Eq.~\eqref{eqn:suzuki-Trotter}.

Let us also consider the analogous quantum evolution in the presence of two separate baths, i.e., without collective effects between the qubits. Then, the master equation driving this dynamics, assuming once again that the frequencies of the qubits and their decay rates are the same, is equivalent to Eq.~\eqref{eqn:LiouvillianSup} without cross terms with $j\neq k$ in the summation (see Appendix~\ref{sec:anSol} and Eq.~\eqref{eqn:liouvLoc} for further details). In this scenario, we have two local Lindblad operators, namely $\sigma_1^-$ and $\sigma_2^-$. Therefore, the multipartite collision model to reproduce the local dynamics requires two ancillas, say respectively qubit $E_1$ and qubit $E_2$, and the same gates as in Eq.~\eqref{eqn:gateSupSub}, both lasting for a time $\Delta t$. Indeed, in the absence of collective effects we do not need to employ the second-order Suzuki-Trotter decomposition as in Eq.~\eqref{eqn:suzuki-Trotter}, while we can just rely on its first-order analogous \cite{Cattaneo2021,Lorenzo2017}.

Let us now fix the values of $\gamma$ and $\Delta t$. We set $\gamma=1$ (after choosing some suitable units of measurement), which is the only physical timescale of the problem. This means that thermalization is reached at a time $t_R\sim O(1)$ \cite{Cattaneo2019b}. The time evolution of the emission power of Eq.~\eqref{eqn:emiPower} is plotted in Fig.~\ref{fig:ideal}(a) for different initial states (superadiant, subradiant, and the {excited} state $\ket{ee}$), and {compared to} the case {of} local dissipation {for an initial} subradiant state. 

{For the quantum simulation, we choose $\Delta t=0.1$ and consider the dynamics until $t=1$, requiring $n=10$ timesteps}. The upper bound on the single-step error, given by Eq.~\eqref{sec:errorBoundTheo}, is plotted in Fig.~\ref{fig:ideal}(b) as a function of the timestep $\Delta t$. These high values of the single-step error are due to the fact that the bound in Eq.~\eqref{eqn:bound} {reflects} the worst-case scenario (the maximization is performed over all the initial states). This is necessary to guarantee the efficiency of the execution of the algorithm, but quite often of little use for practical purposes. Therefore, in Fig.~\ref{fig:ideal}(a) (inset) we plot the trace distance between $\rho_S(t)$, as obtained from Eq.~\eqref{eqn:LiouvillianSup}, and the state simulated by the MCM with $\Delta t=0.1$, as a function of the number of timesteps, {i.e.,}
\begin{equation}
\label{eqn:epsilonIdeal}
    \epsilon_{\mathrm{ideal}}(n)=\normgood{\exp{\mathcal{L}(n\Delta t)}[\rho_S(0)]-(\phi_{\Delta t})^n[\rho_S(0)]}_1,
\end{equation}
for $n=1,\ldots,10$. Note that $\epsilon_{\mathrm{ideal}}(n)$ is different from the global error at the $n$th timestep $\epsilon_g$ introduced in Eq.~\eqref{eqn:global11}. Indeed, in Eq.~\eqref{eqn:epsilonIdeal} there is no maximization over all the possible initial states, {since we focus on a specific} initial state. As a consequence, $\epsilon_{\mathrm{ideal}}(n)$ is much lower than $\epsilon_g$. In fact, in Fig.~\ref{fig:ideal}(a) (inset) we observe that, even with a large timestep such as $\Delta t=0.1$, the MCM is able to simulate the state of the dynamics at time $t$ with high accuracy.

\section{Error bound for the noisy simulation}
\label{sec:errorBoundTheo}

In Ref.~\cite{Cattaneo2021} an error bound for the ideal case of the MCM based on the $1\rightarrow 1$ superoperator norm is computed. For a single step of the MCM, it is expressed as in Eq.~\eqref{eqn:bound}. Here, {we} estimate an analogous upper bound for {the case of noisy gates}. {We use} the diamond norm instead of the $1\rightarrow 1$ superoperator norm to obtain bounds expressed in terms of error values that can be employed to guarantee fault-tolerant quantum computation. 

Let us first replace the ideal quantum map $\phi_{\Delta t}$, defined in Eq.~\eqref{eqn:quantumMap}, with a noisy one, which we term as $\phi_{\Delta t}^*$. The error we want to estimate is:
\begin{equation}
\label{eqn:globalErrorNoisy}
    \epsilon_g^*=\normgood{(\phi_{\Delta t}^*)^n-\exp \mathcal{L}t}_\Diamond, \text{ with }t=n\Delta t.
\end{equation}
Thanks to the triangle inequality, we have:
\begin{equation}
\begin{split}
\label{eqn:globalErrorNoisyBound}
    \epsilon_g^*\leq &\normgood{(\phi_{\Delta t})^n-\exp \mathcal{L}t}_\Diamond+\normgood{(\phi_{\Delta t}^*)^n-(\phi_{\Delta t})^n}_\Diamond\\
    &\leq n \left(\epsilon_s^{\Diamond}+\epsilon_{m}^*\right),
\end{split}
\end{equation}
where we have introduced the errors
\begin{equation}
\label{eqn:normsNew}
\epsilon_s^{\Diamond}=\normgood{\phi_{\Delta t}-\exp \mathcal{L}\Delta t}_\Diamond,\quad \epsilon_{m}^*=\normgood{\phi_{\Delta t}^*-\phi_{\Delta t}}_\Diamond.
\end{equation}
Let us focus on these norms for a single application of the MCM. 
We first address the bound for the ideal case based on the diamond norm $\epsilon_s^{\Diamond}$, and then the diamond norm $\epsilon_{m}^*$ between the ideal and noisy MCM maps. We give here only the final result, while the derivation of the bounds can be found in Appendix~\ref{sec:proofs}. Our first statement is:
\begin{prop}
The ideal error bound $\mathcal{B}_{1\rightarrow 1}$ evaluated through the $1\rightarrow 1$ superoperator norm as in Eq.~\eqref{eqn:bound} is valid also for the diamond norm:
\begin{equation}
    \label{eqn:boundDiamondIdeal}
\epsilon_s^{\Diamond}\leq\mathcal{B}_{1\rightarrow 1},
\end{equation}
where the expression $\mathcal{B}_{1\rightarrow 1}$ was introduced in Eq.~\eqref{eqn:bound}.
\end{prop}
This means that all the results of Ref.~\cite{Cattaneo2021} about the scaling of the error can be trivially extended to a scenario where the diamond norm is employed. In particular, the efficient quantum simulation of the MCM is guaranteed also through the diamond norm. Note that, for instance, the behavior of the upper bound shown in Fig.~\ref{fig:ideal}(b) holds also for $\epsilon_s^\Diamond$.

Let us now focus on $\epsilon_m^*$. Our aim is to estimate ``how far'' the noisy implementation of the MCM is from its ideal analog. To do so, we may assume that each ancillary qubit is not prepared in the ideal initial state $\rho_{E,i}$, but in the noisy $\rho_{E,i}^*=\mathcal{G}_i[\rho_{E,i}]$, where $\mathcal{G}_i$ is a {known quantum channel characterizing the noise on the state preparation of the $i$th ancilla in the actual device.} The quantum channel for the total unitary evolution of the system+ancillary qubits of the ideal MCM for a single timestep $\Delta t$ is $\mathcal{U}_{sim}(\Delta t)=U_{sim}(\Delta t) \cdot U^\dagger_{sim}(\Delta t)$ (see Eq.~\eqref{eqn:evolution_sim}). The unitary evolution $U_{sim}(\Delta t)$ can also be decomposed as a composition of many quantum gates on a quantum computer, such as $\mathcal{U}_{sim}(\Delta t)=\prod_j \mathcal{U}_j$. These gates typically act on both the system and ancillary qubits. Let us now suppose that, on a real platform, each of these gates is not ideal but noisy, and can be represented by the quantum channel $\mathcal{E}_j$. Then, we find:
\begin{prop}
\begin{equation}
\begin{split}
\label{eqn:boundNoisyMCM}
\epsilon_m^*&\leq\sum_j\normgood{\mathcal{E}_j-\mathcal{U}_j}_\Diamond+\sum_i\normgood{\mathcal{G}_i-\mar{\mathcal{I}}_{E}}_\Diamond\\
&=2\left(\sum_j d_\Diamond(\mathcal{U}_j,\mathcal{E}_j)+\sum_id_\Diamond(\mar{\mathcal{I}}_{E},\mathcal{G}_i)\right),
\end{split}
\end{equation}
where $\mathcal{G}_i$ and $\mathcal{E}_j$ are the noisy channels introduced above, while the diamond distance $d_\Diamond$ is defined in Eq.~\eqref{eqn:diamondDistance}.
\end{prop}
    That is to say, even though the MCM map acts on the state of the system only, we can estimate an upper error bound for the noisy map that is equal to the sum of the individual errors for each quantum gate between the system qubits and the ancillas (including the preparation of the initial states of the ancillary qubits). Note that the above error bound is valid also for modified versions of the MCM, where, for instance, the ancillary qubits can be prepared in an initial entangled state. Moreover, the above bound is robust even under more general sources of error, e.g., if the initial state of system qubits+ancillas is accidentally entangled. Indeed, the diamond distance $\normgood{\mathcal{E}_j-\mathcal{U}_j}_\Diamond$ involves a maximization over all the possible initial states of the overall system\footnote{That is, including system qubits + ancillary qubits of the MCM + a copy of the total (system+ancillas of the MCM) Hilbert space according to Eq.~\eqref{eqn:diamondNorm}, i.e., $\mathcal{H}_S\otimes\mathcal{H}_E\otimes\mathcal{H}_A\otimes\mathcal{H}_B$ using the labeling we introduce in Appendix~\ref{sec:proofs}.}, including the entangled ones. In such a scenario, the noisy preparation given by the set of $\mathcal{G}_i$ can then be extended to act on both system qubits and ancillas, yielding an entangled state instead of an initial state that is separable between system and ancillary qubits.

To summarize, the results stated in Proposition 1 and Proposition 2 can be employed to estimate a general upper bound for the global error of the noisy MCM map, according to Eq.~\eqref{eqn:globalErrorNoisyBound}. More specifically, Eq.~\eqref{eqn:boundDiamondIdeal} expresses the error due to the finite (and not infinitesimal) timestep $\Delta t$ in the ideal algorithm. Clearly, the lower $\Delta t$ the better, as shown in Fig.~\ref{fig:ideal}(b). Instead, Eq.~\eqref{eqn:boundNoisyMCM} states that the error due to a noisy MCM protocol can be decomposed into the sum of the individual errors for each quantum gate employed in the algorithm, including the state preparation of the ancillary qubits. The diamond distances in Eq.~\eqref{eqn:boundNoisyMCM} can be estimated experimentally, and we will show their values for some CNOT gates employed to simulate the MCM on the IBM quantum computers (see the discussion in Sec.~\ref{sec:IBM} and Appendix~\ref{sec:diamond}).

We stress again that we are using the diamond distance, which is employed to prove the formal theorems about fault-tolerant quantum computation. This means that the distances between noisy and ideal gates that appear in Eq.~\eqref{eqn:boundNoisyMCM} express exactly the errors that we must keep below a certain threshold to guarantee fault-tolerant computation. Moreover, recall that the diamond norm addresses the worst-case scenario, so the upper error bound can be higher than the {actual error} in a single implementation of the algorithm (compare Figs.~\ref{fig:ideal}(a) and (b)).

\section{Experimental demonstration on IBM quantum computers}
\label{sec:IBM}

\subsection{Introduction to the experimental results}
In this section, we present the implementation of the multipartite collision model on {a near-term superconducting IBM quantum computer available on the cloud, programmed through the library Qiskit \cite{Qiskit}.  In particular, we will show some experimental results obtained on the limited-access 16-qubit \texttt{ibmq\_guadalupe}, while more results on the 27-qubit \texttt{ibmq\_toronto} are discussed in Appendix~\ref{sec:toronto}}. 

Our aim is twofold. On the one hand, we want to show how even current near-term devices can display non-trivial dissipative collective effects in a quantum simulation by means of the multipartite collision model. On the other hand, we intend to investigate the properties of noise in these platforms via process tomography, and to relate them with the accuracy of the simulation of the MCM. By doing this, we aim to investigate whether the quantum physicists who make use of near-term quantum computers on the cloud can infer the accuracy of the simulations they would like to run without having access to the hardware. We will see that gate process tomography can provide us with useful information about the precision of the quantum simulation despite its drawbacks that are discussed in the literature \cite{Merkel2013,Nielsen2021}, namely its sensitivity to preparation and measurement (SPAM) errors.

In this paper we present two different sets of results. The first one (discussed in the present section) shows clear signatures of collective effects, and the accuracy of the simulation is reasonably good. The second one is presented in Appendix~\ref{sec:detrimental} and displays noisier results, and the simulated MCM yields results that are quite far from their {expected} values. We will show, however, that this can be traced back to some highly noisy gates that are repetitively employed in the algorithm, the noise properties of which are captured by our analysis based on process tomography. Whether these high levels of noise are due to an incorrect application of the quantum gate itself or to SPAM errors may be a matter of debate. However, their signatures are clearly evident in the results of the quantum computation, which, ultimately, is what we are interested in when we run algorithms for quantum simulation on near-term quantum computers. 

The experimental details and the platform schemes can be found in Appendix~\ref{sec:expScheme}. Shortly, the protocols that, for the above-mentioned purposes, we have to run on the devices are (note that process tomography is not needed for the MCM, but only to assess errors):
\begin{enumerate}
    \item Algorithm implementing the MCM until the step $n=5$ for the three different initial states explored in Sec.~\ref{sec:subSup} and for the local dynamics, {and performing measurements} in the {computational} basis.
    \item State tomography of the system at each timestep of the algorithm.
    \item Process tomography \cite{nielsenchuang} of the CNOT gates employed in the algorithm, in order to estimate the noise properties. We do not address the properties of the single-qubit gates, because their error is usually a couple of orders of magnitude lower than the one of two-qubit gates (see Appendix~\ref{sec:expScheme} for further details). 
    \item Readout error mitigation on the qubits {measured in both} the algorithm and the gate process tomography. This is a standard procedure that can be implemented on the IBM near-term devices through the Qiskit library \cite{Qiskit}. It detects possible systematic errors in the outcomes of the measurements on the set of qubits of interest, and allows for readout error mitigation. All the outcomes we will show in the manuscript have been obtained after applying the proper readout error mitigation procedure.
\end{enumerate}

Note that we will simulate $n=5$ steps of the MCM, using the same model parameters that have been employed in Sec.~\ref{sec:subSup} (namely, decay rate $\gamma=1$, timestep $\Delta t=0.1$). The specifics of the near-term devices we are employing and the noise level do not allow for more collisions between the system qubits and the ancillas. In any case, we will see that 5 steps are already enough to observe collective effects in the dynamics.

{It is worth stressing that a crucial challenge in the simulation of the MCM is getting a new ancilla initialized in the ground state after each collision, as required by the steps of the algorithm discussed in Appendix~\ref{sec:steps}. In general, in the topology of a quantum computer only one or two ancillary qubits are directly connected to (i.e., can perform operations with)  both the system qubits. Therefore, we have generated a new refreshed ancilla after each timestep by swapping the state of the common ancillary qubit with the one of the nearby qubits prepared in the ground state. To do this, we need more and more swap gates as the number of collisions increases, giving rise to a ``train of ancillas'' that must be subsequently swapped to get to interact with the system qubits. More details can be found in Appendix~\ref{sec:expScheme}. A different solution may consist in employing a reset gate to reinitialize the state of the common ancillary qubit in the ground state after each collision. However, this did not work on the platforms we have used due to decoherence effects, as shown in Appendix~\ref{sec:reset}.}

\begin{figure*}
\centering
  \includegraphics[width=.75\textwidth]{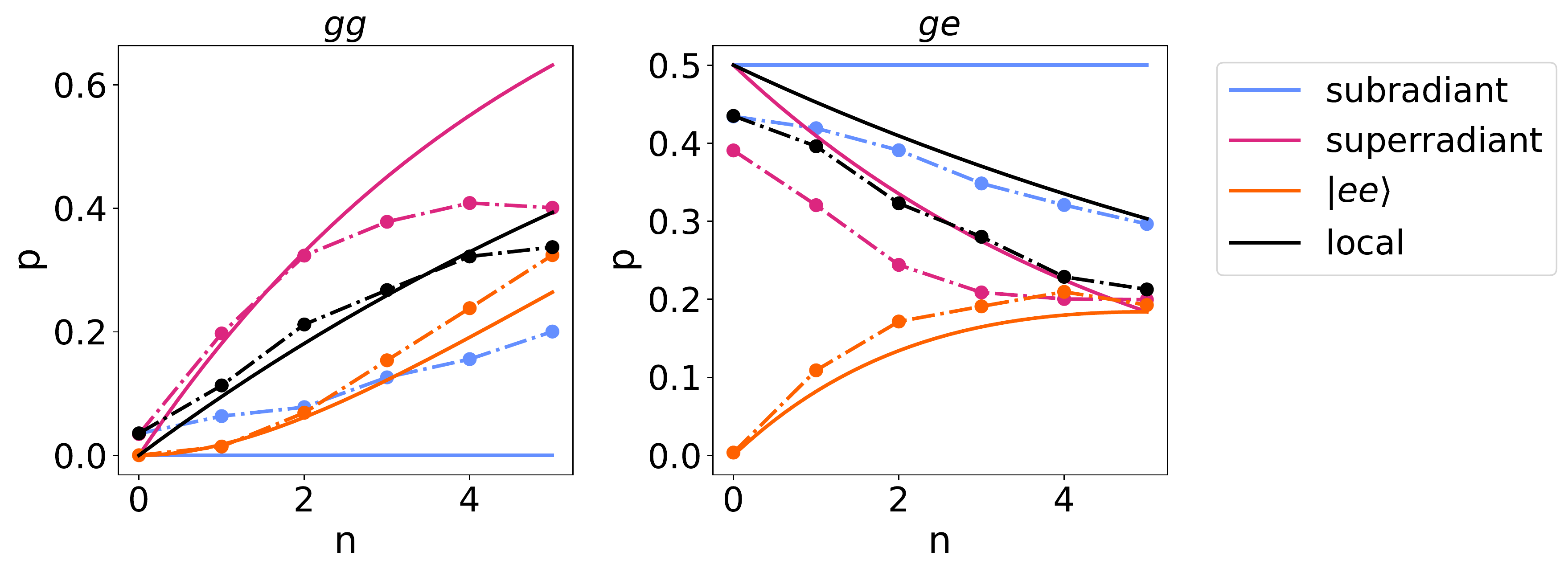}\\
  \includegraphics[width=.75\textwidth]{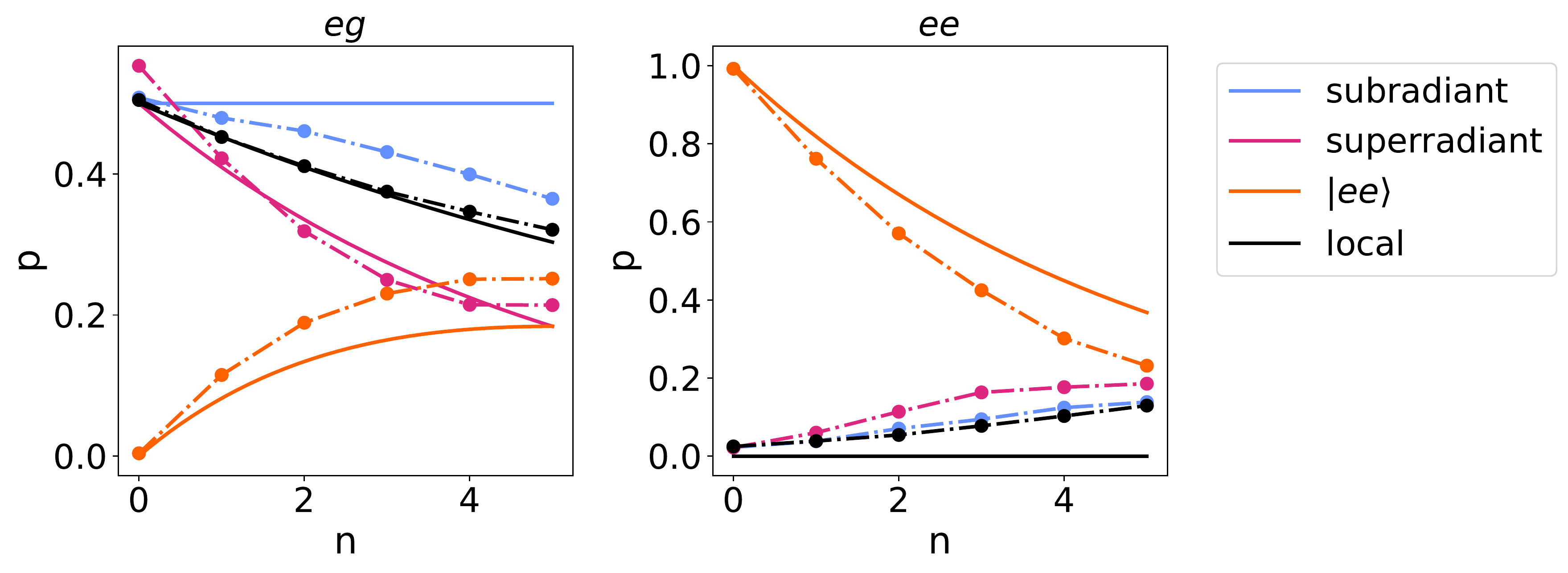}%
    \caption{Probability of finding the states $\ket{gg}$, $\ket{ge}$, $\ket{eg}$, $\ket{ee}$ through a projective measurement in the {computational} basis as a function of the number of steps in the MCM (first set of results), when the initial state is the subradiant state (light blue), superradiant state (magenta), $\ket{ee}$ (orange), and when the dynamics is local starting from the subradiant state (black). Solid lines: theoretical prediction based on the master equation. Markers of the dash-dotted lines: experimental values. The results have been obtained as averages over 37 realizations of the protocol, and the error bars are within the markers.}
    \label{fig:countsOne}
    \end{figure*}
    
    \begin{figure*}
  \includegraphics[scale=0.32]{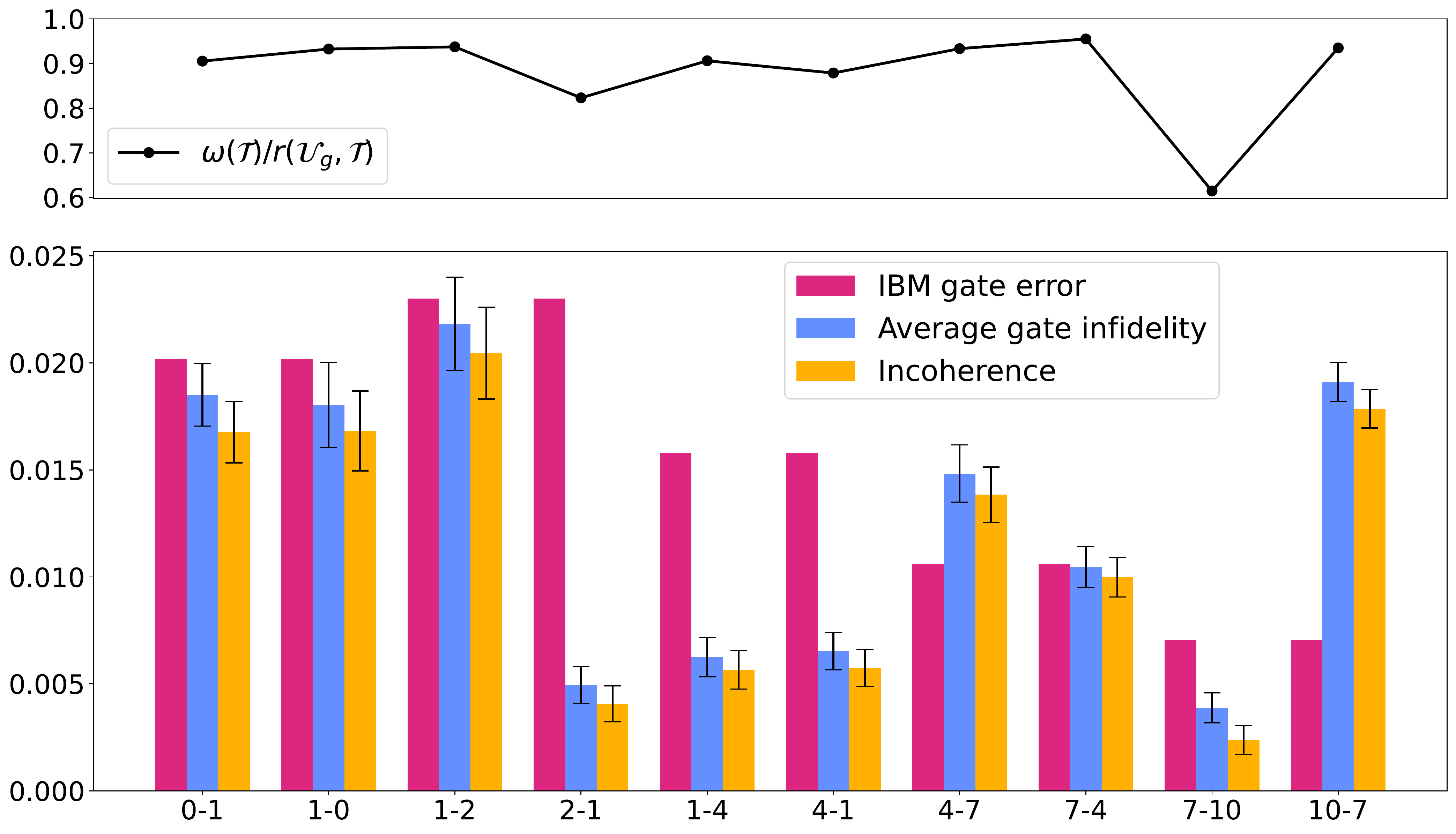}%
    \caption{Error analysis of the set of CNOT gates in \texttt{ibmq\_guadalupe} employed for the simulation of the first set of results in Fig.~\ref{fig:countsOne}. The tick 0--1 on the x axis corresponds to the CNOT gate where ``qubit 0'' in \texttt{ibmq\_guadalupe} is the control qubit and ``qubit 1'' is the target. Lower plot: gate error provide by IBM (magenta), experimental average gate infidelity $r(\mathcal{U}_g,\mathcal{T})$ via full process tomography (light blue), and experimental incoherence $\omega(\mathcal{T})$ (yellow), as defined in Eq.~\eqref{eqn:incoherence}. The error bars on the values of $r(\mathcal{U}_g,\mathcal{T})$ and $\omega(\mathcal{T})$ are the standard deviations of 100 realizations of a random sample of the experimental data via bootstrapping. Upper plot: ratio between the incoherence and the experimental average gate infidelity of each CNOT gate.}
    \label{fig:erroran1}
\end{figure*}

\subsection{Experimental MCM outcomes}

A first set of results and measurement outcomes in the {computational} basis of the quantum computer is shown in Fig.~\ref{fig:countsOne} as a function of the number of steps in the MCM, from $n=0$ (state preparation) to $n=5$. {The {computational} basis ``00'', ``01'', ``10'', ``11'' on the IBM quantum computer \cite{Qiskit} corresponds to the physical basis $\ket{gg}$, $\ket{ge}$, $\ket{eg}$, $\ket{ee}$ of the two qubits.}

The results display a clear evidence of dissipative collective effects in the system dynamics. To see this, let us compare the behavior of the subradiant and superradiant evolution with the one of the local dynamics. The ideal local decay of the subradiant state (solid black lines) can be considered as a benchmark to discriminate between collective and non-collective dynamics. Indeed, as shown also in Fig.~\ref{fig:ideal}(a), it corresponds to an exponential decay driven by the local dissipation rate $\gamma$ (see Appendix~\ref{sec:anSol} for the analytical solution of the dynamics). Then, the emergence of collective effects is detected by either the presence of a slower decaying eigenspace of the dynamics (``subradiance'') or by a much faster decay (``superradiance''), depending on the initial state of the system. This is exactly what we observe in the experimental data depicted in Fig.~\ref{fig:countsOne}: if we start in the subradiant state (dash-dotted light blue lines), the decay of the excited population levels ($\ket{ge}$ and $\ket{eg}$) into the ground state ($\ket{gg}$) is slower than the one given by the ideal local dynamics, i.e., they are decaying with a rate that is smaller than $\gamma$. This can be possible only in the presence of collective effects between the two qubits. Note that, although the light blue markers are below the black solid line in the {evolution of $\ket{ge}$}, this is not due to a faster, local decay of the subradiant state, but to an imprecise state preparation at time $t=0$ for which the initial population of $\ket{ge}$ is lower than 0.5. Its decay, however, is again slower than $e^{-\gamma t}$. The same considerations apply for the evolution of the superradiant state (dash-dotted magenta lines): the decay of the initially excited populations ($\ket{ge}$ and $\ket{eg}$) into the ground state ($\ket{gg}$) is clearly faster than the local decay and, at least during the first timesteps of the dynamics, it follows the two times faster decay driven by $e^{-2\gamma t}$ (solid magenta lines). The experimental local decay of the subradiant state (black markers) approximately follows the ideal local dynamics, and it decays as $e^{-\gamma t}$, as we were expecting. Finally, the collective evolution of the state $\ket{ee}$ (orange markers) is a linear combination of subradiant and superradiant dynamics (see Appendix~\ref{sec:anSol}), and it well captures its ideal behavior (orange solide lines). Moreover, it does not run into a large error in the state preparation, since preparing $\ket{ee}$ (starting from $\ket{gg}$) requires only single-qubit gates. In contrast, the initialization of the super- and sub-radiant states involves two-qubit gates, which are much noisier than single-qubit operations, and this is why their state preparation is more imprecise.

It is worth noting, however, that the subradiant state is decaying, while it should be stationary according to the master equation~\eqref{eqn:LiouvillianSup}. In other words, there are leakages out of the decoherence-free subspace. This is due to the errors in the gates we employed in the algorithm, which are inevitably noisy in near-term devices. {Another source of leakages is due to the finite Trotter timestep $\Delta t$ of the algorithm. However, as one can infer from the inset of Fig.~\ref{fig:ideal}(a), the discrepancy between the true dynamics and the ideal simulation is very small for the subradiant state, therefore this source of leakages is basically negligible in the present scenario. The interested reader can check in Appendix~\ref{sec:trace} a quantitative comparison between the error in the experimental results of Fig.~\ref{fig:countsOne} due to noisy gates and the ideal error of the algorithm. For the subradiant state, the ratio between ideal and experimental error is always smaller than 1\%.} 

{We stress that, despite the dissipation due to the gate errors}, we have proven that the multipartite collision model produces a more robust subradiant state than in the presence of local decay only. Therefore, our algorithm is able to preserve the populations of an entangled state of the qubits for a longer time than in the absence of collective effects, even on near-term devices with a considerable level of decoherence.

\begin{figure*}
    \centering
  \includegraphics[scale=0.38]{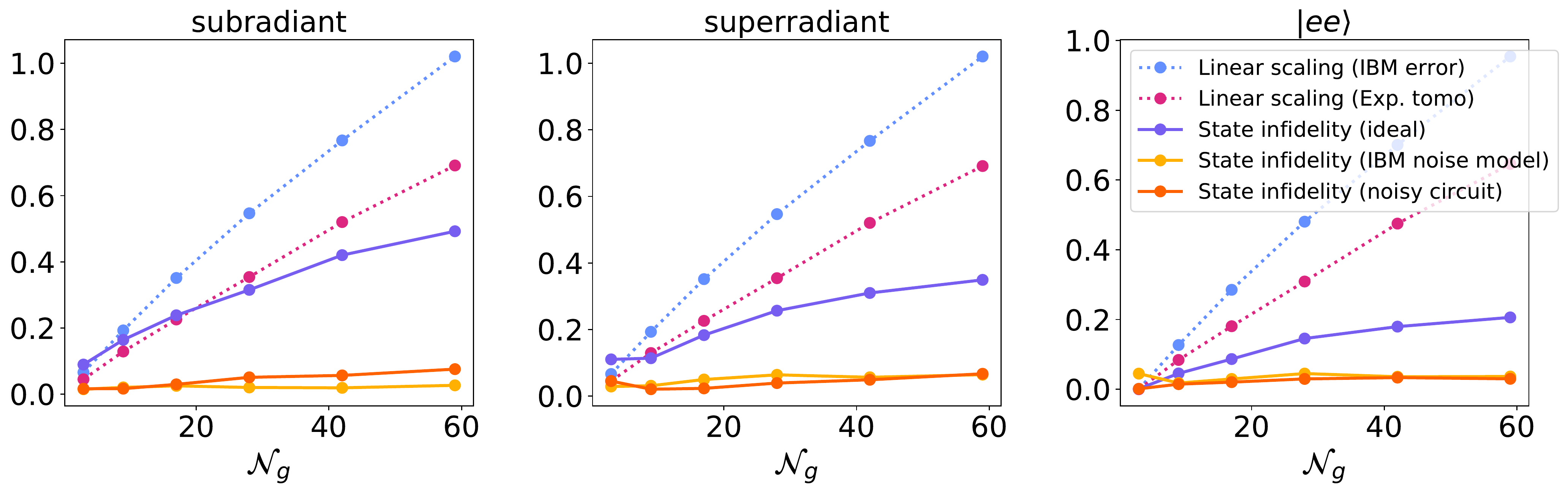}%
    \caption{Scaling of the error as a function of the number of CNOT gates in the algorithm $\mathcal{N}_g$ during the simulation of the collective dynamics displayed in Fig.~\ref{fig:countsOne}. The markers indicate the steps of the collision model. Dotted light blue line and dotted magenta line: linear scaling of respectively the IBM gate error and the experimental average gate infidelity, given in Fig.~\ref{fig:erroran1}. Solid violet line: infidelity between the experimental state and the ideal one. The errors bars for the latter quantity have been obtained as the standard deviations of 100 realizations of a random sample of the state tomography data via bootstrapping, and they are within the markers. Solid orange line: infidelity between the experimental state and the state simulated on the noisy circuit. Solid yellow line: infidelity between the experimental state and the state simulated via the IBM noise model.}
    \label{fig:errortimestep1}
\end{figure*}

\begin{figure*}
    \centering
  \includegraphics[scale=0.38]{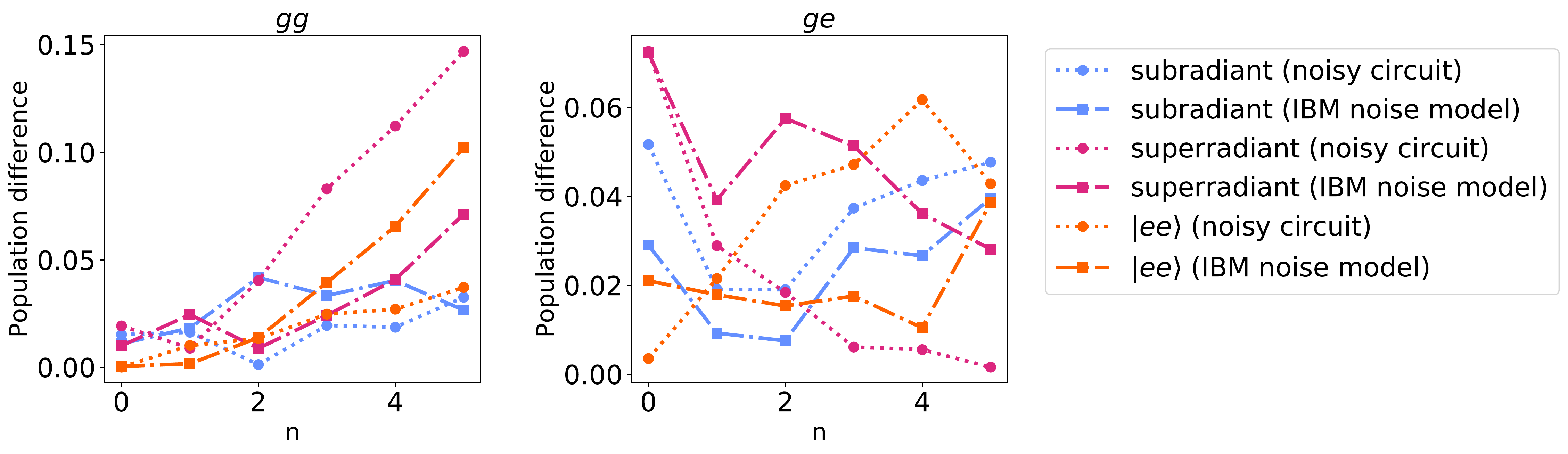}\\
  \includegraphics[scale=0.38]{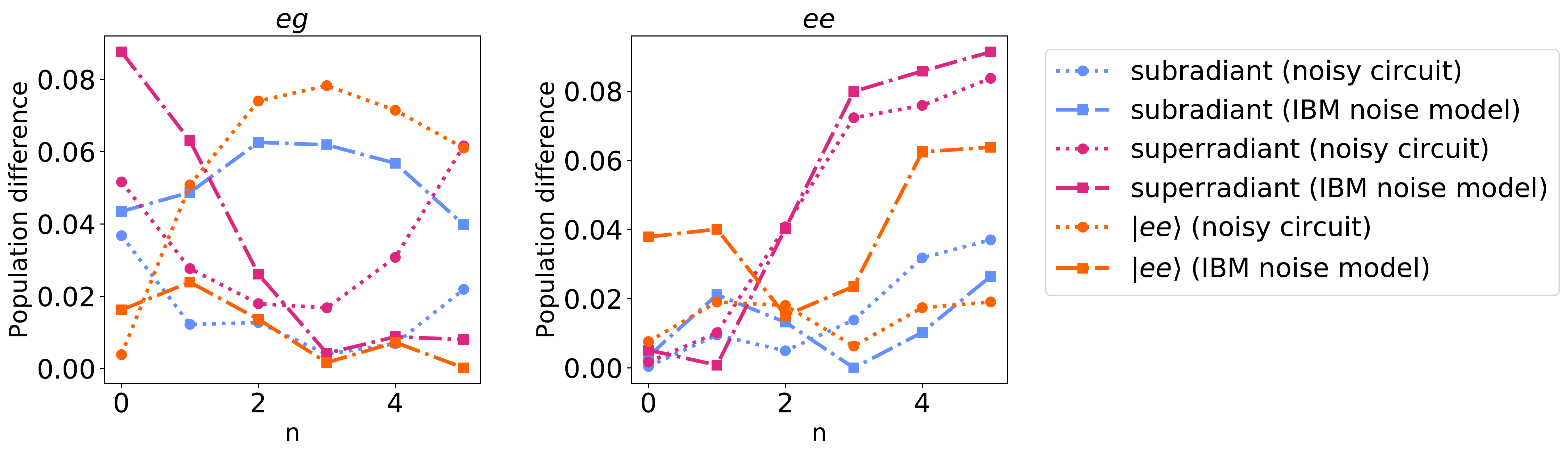}%
    \caption{Absolute value of the population difference between the experimental state and the state simulated through the IBM noise model (squared markers of the dash-dotted lines) and between the experimental state and the  state simulated on the noisy circuit (round markers of the dotted lines) as a function of the MCM timesteps.}
    \label{fig:errormismatch1}
\end{figure*}

\subsection{Gate error analysis}
\label{sec:gateErrorAn}
Let us now focus on the gate analysis displayed in Fig.~\ref{fig:erroran1}. We compare the experimental average gate infidelity of the  CNOTs in the algorithm, computed by reconstructing the Choi matrix of the gate \cite{Choi1975} through gate process tomography, with the gate error provided by IBM and extrapolated through a randomized benchmarking protocol \cite{Magesan2011a,Qiskit}. We also plot the value of the incoherence $\omega(\mathcal{T})$ given by Eq.~\eqref{eqn:incoherence}. The experimental values have been resampled 100 times via bootstrapping to estimate the standard deviations of the samples (depicted in Fig.~\ref{fig:erroran1} as error bars). We have also computed the values of the diamond distances between ideal and experimental gates, and a discussion about them can be found in Appendix~\ref{sec:diamond}. 

Note that the CNOTs between qubit ``10'' and ``12'' are not taken into account due to an error in the experimental outputs. However, these gates are employed only a single time during the whole protocol to switch the fourth and fifth collision ancillas, so their features are of little interest for our analysis. 

We first make two major remarks:
\begin{enumerate}[label=(\roman*)]
    \item As shown in the upper plot of Fig.~\ref{fig:erroran1}, the ratio between the incoherence and the average gate infidelity is larger than 0.9 for almost all gates. This means that the ``coherence of noise'' \cite{Wallman2015a} in the hardware is usually quite low, and the dominant source of error is incoherent. This agrees with previous studies on the noise properties of the gate pulses on the IBM quantum computers \cite{Willsch2017}.
    \item The experimental average gate infidelity is of the same order of magnitude as the IBM gate error, but it can quite remarkably differ from the latter (e.g., see the CNOT 2--1 or 1--4). So, we are observing a discrepancy between the average gate fidelity via process tomography and the value provided by IBM, which is typically obtained through randomized benchmarking. Similar remarks on the IBM quantum devices can be found in the literature \cite{Michielsen2017,Bultrini2021}. For instance, the full process tomography shows how the average gate infidelity of a CNOT gate between the same pair of qubits can differ when we switch the control and target qubits, while this is not captured by the IBM error.
\end{enumerate}
 Overall, the discrepancy between the experimental average gate infidelity and the IBM error may be explained as follows: 1) The IBM error is describing the average infidelity of a noise channel obtained as, following standard randomized benchmarking, the average channel between the twirling gates of the Clifford group (see, e.g., Refs.~\cite{Magesan2012a,Wallman2016,Merkel2021} for more rigorous details); this is providing us with a useful measure to benchmark the average noise on the selected pair of qubits, but it is different from the average gate infidelity of a specific CNOT gate. 2) Even if readout error mitigation has been applied, some residual SPAM errors might be present in the characterization of the average gate infidelities, while the IBM gate error is SPAM-free.

Even if we cannot rule out the presence of SPAM errors in the values of the average gate infidelities (this is a well-known drawback of process tomography \cite{Merkel2013}), we point out that these errors have been mitigated through the procedure for readout error mitigation. Indeed, the experimental values of the average gate infidelities without error mitigation are huge (typically around 5 times larger than the values we plotted in Fig.~\ref{fig:erroran1}). Moreover, the gates for state preparation and state measurement in the circuits for the full process tomography are only single-qubit gates, whose gate error is typically two orders of magnitude smaller than the two-qubit gate error \cite{Qiskit}; for instance, the state preparation of $\ket{ee}$, which is the only one not involving two-qubit gates, is much better than the preparation of the subradiant or superradiant states. So, it may be reasonable to posit that the state preparation error in the characterization of the average gate infidelities is not as relevant as the error due to the application of the gate itself. 


\subsection{Scaling of the experimental infidelity}
\label{sec:linearscaling}

After computing the Choi matrices of the quantum gates we employ in the experiment via full process tomography and the figures of merit depicted in Fig.~\ref{fig:erroran1}, we can try to employ them to better understand the noise we will face during an execution of the algorithm. A possible way to do so is to simulate a noisy version of the algorithm run on a ``noisy circuit'', which we construct by replacing all the CNOT gates of the ideal circuit for the MCM with their corresponding non-ideal quantum channels described by the Choi matrices we have estimated experimentally. Then, we can compare the state simulated on this noisy circuit with the experimental state that we have reconstructed through state tomography (the data for the state tomography have been resampled 100 times through bootstrapping and the error bars are within the markers in Fig.~\ref{fig:errortimestep1}). Yet another noise model we may test is the standard noise model from the backend provided by IBM \cite{Qiskit}. This is built as follows: i) for each single-qubit gate we add a depolarizing channel with error rate equal to the IBM gate error, followed by a thermal relaxation through pure dephasing \cite{BreuerPetruccione} with rate $T_2$ and through pure dissipation with rate $T_1$ (these values are available on the IBM website and typically, for \texttt{ibmq\_guadalupe}, they are of the order of 100 $\mu$s, while the qubit frequencies are of the order of $2\pi\times 5$GHz) ii) for each two-qubit gate we apply a two-qubit depolarizing channel with error rate equal to the IBM gate error given in Fig.~\ref{fig:erroran1}, followed by a local thermal relaxation as for single-qubit gates. The IBM noise model takes into account single-qubit gate errors and the ``natural'' dissipation acting on each qubit, while the noise model based on the noisy circuit we reconstructed through the experimental Choi matrices focuses on the two-qubit gate errors only.

In Fig.~\ref{fig:errortimestep1} we compare the states simulated according to the noise models presented above with the ideal and experimental states. The solid violet line depicts the infidelity between the ideal and experimental state as a function of the number of CNOT gates in the protocol $\mathcal{N}_g$. The markers indicate the state at each step of the algorithm, from $n=0$ (state preparation) to $n=5$. Note that the number of gates to implement the $n$th step is higher than for the step $n-1$, as explained in Appendix~\ref{sec:expScheme}. We can compare its scaling with the linear sum (as a function of $\mathcal{N}_g$) of the average gate infidelities of all the CNOTs employed in the algorithm, taking into account their repetitions, and following the order in which they appear in the circuit (i.e., $\mathcal{N}_g=1$ on the $x$ axis denotes the first gate we employ in the algorithm, and so on). We plot this linear scaling for both the experimental average gate infidelity obtained through gate process tomography and the IBM gate error, which are displayed in Fig.~\ref{fig:erroran1}. Note that these curves are not exactly straight lines, because the average gate fidelities of the gates employed in the algorithm are in general different.
According to the discussion in Sec.~\ref{sec:comparingFig}, we may expect the experimental state infidelity to scale linearly when the noise source is mostly incoherent \cite{Carignan-Dugas2019a,Kueng2016}, at least during the first steps of the algorithm (as soon as the number of CNOT gates $\mathcal{N}_g$ increases, it is reasonable to assume that the experimental errors will drive the state of the system towards some sort of mixed state, so that the experimental infidelity will saturate at a value that is not captured by the linear scaling in Fig.~\ref{fig:errortimestep1}). We indeed observe a linear scaling, or also sublinear, as a function of the number of gates. The linear sum of the experimental average gate infidelities captures the first stages of the algorithm, but then overestimates the error, as we may expect due to some sort of saturation toward a mixed state. This may suggest that, in the platform we are considering, the average gate fidelity of composed channels \cite{Carignan-Dugas2019a} scales in a favorable way, which may also be sublinear. The scaling of the trace distance between experimental and ideal state as a function of $\mathcal{N}_g$ is also linear or sublinear (we refer the interested reader to the discussion in Appendix~\ref{sec:trace}).

Let us now focus on the solid orange and yellow lines, which depict the infidelity between the experimental state and respectively the state simulated through the noisy circuit based on experimental Choi matrices and the state simulated through the IBM noise model. Remarkably, both error models provide us with a good prediction for the experimental state, as the infidelity between the latter and the simulated states never exceeds 0.1, and is usually lower than 0.05. For the noisy circuit noise model, the discrepancy between the experimental and simulated states may be explained through the presence of crosstalks or correlated measurement errors on the backend, i.e., the quantum channel associated with a gate when operated on its own may be different from the corresponding quantum channel when the same gate is applied during a more complex algorithm with multiple quantum operations on different qubits at the same time. Another reason for this discrepancy may be residual SPAM errors in the characterization of the Choi matrices of the gates.

Interestingly, both noise models have a quite similar performance, although the noise model based on experimental Choi matrices may outperform the IBM noise model in the presence of very noisy gates (see the second set of results in Appendix~\ref{sec:detrimental}). This means that a noise model that only addresses two-qubit errors can be as good as a more complete noise model if the CNOT errors are properly characterized, highlighting how the latter are the main source of noise in near-term computers. Finally, note that, despite the two noise models having a similar performance, as captured by the state infidelity between experimental and simulated states, the states they simulate may actually be quite different. Indeed, Fig.~\ref{fig:errormismatch1} shows the absolute value of the population difference between the experimental (results in Fig.~\ref{fig:countsOne}) and simulated states for both noise models. We observe that one model outperforms the other for some timesteps and for different observables, but overall they display a similar performance, despite predicting quite different quantum states. A possible improvement of our noise model may consist in including a depolarizing channel for each single-qubit gate, where the decay rate of the channel is obtained through randomized benchmarking (e.g., the IBM-provided error values can be used for this). We leave this possibility for future works.

\section{Concluding remarks and perspectives}
\label{sec:conclusions}
We have presented the first fully quantum digital simulation of dissipative collective effects on a quantum computer, and we have analyzed both theoretically and experimentally the impact of noisy gates on the algorithm. 

{State-of-the-art universal quantum computers do not allow for fault-tolerant computation yet, while they are subject to a considerable level of noise. However, it is remarkable} that the algorithm we employed, namely the \textit{multipartite collision model} (MCM) \cite{Cattaneo2021}, simulates the superradiant and subradiant dynamics of two qubits colliding with a common ancilla with a good degree of accuracy (Fig.~\ref{fig:countsOne} and Fig.~\ref{fig:countsTwo} in Appendix~\ref{sec:further}){, and that a noise model based on our experimental noise analysis is able to estimate with reasonable precision the distance between the ideal and experimental states (Fig.~\ref{fig:errortimestep1} and Fig.~\ref{fig:errortimestep2} in Appendix~\ref{sec:detrimental})}.

{More specifically}, we have {simulated on a quantum computer} the enhanced decay of the two-qubit state prepared in $(\ket{eg}+\ket{ge})/\sqrt{2}$, which is a paradigmatic signature of superradiance, and the very slow decay of the state $(\ket{eg}-\ket{ge})/\sqrt{2}$, which displays the emergence of subradiance. As a benchmark for these collective phenomena, we have also simulated the two-qubit dynamics in the presence of local decay only, and the results follow the theoretical scaling.


{We have stressed the importance of the rigorous analysis of the gate errors when dealing with current quantum devices, and here we have addressed this issue also theoretically}.
In Ref.~\cite{Cattaneo2021}, an error bound for the ideal multipartite collision model was derived using the $1\rightarrow 1$ superoperator norm. This error is essentially due to the inevitable choice of a small but finite algorithm timestep $\Delta t$, while the second-order Suzuki-Trotter decomposition of the MCM simulates the exact dynamics in the limit of infinitesimal timestep only.
In Sec.~\ref{sec:errorBoundTheo}, we have presented a refined error bound that takes into account the possibility of employing noisy gates in the practical implementation of the MCM. Specifically, Proposition 1 generalizes the bound of Ref.~\cite{Cattaneo2021} by making use of the diamond norm, which is a more precise norm for the distance between quantum channels and, crucially, is employed in the estimation of a rigorous error threshold for quantum fault-tolerant computation. Moreover, Proposition 2 estimates an upper bound based on the diamond norm for the error we are incurring by using imperfect gates. While the quantum map associated with the MCM acts on the state of the system only, Proposition 2 expresses the error bound by taking into account the action of each gate on both the system and the ancillary qubits, and decomposes it into the sum of the individual errors of each quantum gate. 

On the experimental side, the results of the noise analysis are depicted in Figs.~\ref{fig:erroran1} and~\ref{fig:erroran2} in Appendix~\ref{sec:detrimental} for the average gate infidelity and the incoherence, while Fig.~\ref{fig:erroranDiam1} in Appendix~\ref{sec:diamond} shows the experimental diamond distance between the employed CNOTs and their ideal counterparts. To obtain these figures of merit, we have performed the full process tomography of all the CNOT gates employed in the algorithm. To reduce the SPAM errors that are the major drawback of two-qubit process tomography, we have applied readout error mitigation. However, the reader must be aware that residual SPAM errors may be present in the results of our experimental noise analysis. Then, we have compared these results with the gate errors provided by IBM, which are obtained through randomized benchmarking.

Our findings indicate that the ratio between incoherence and average gate infidelity is almost always larger than 0.9, therefore the major source of error is dissipative, i.e., the ``coherence of noise'' \cite{Wallman2015a} is low. We have also observed that the experimental average gate infidelity computed through full process tomography can sometimes remarkably differ from the IBM gate error (see Figs.~\ref{fig:erroran1} and~\ref{fig:erroran2} in Appendix~\ref{sec:detrimental}). Moreover, the scaling of the infidelity between exact and simulated states as a function of the number of gates is linear or sublinear (see Figs.~\ref{fig:errortimestep1} and~\ref{fig:errortimestep2}), i.e., much better than the worst-case scenario with quadratic scaling. This may be due to the fact that the major source of errors is dissipative and not coherent. Finally, the diamond errors of the CNOT gates we have obtained are of the order of $5\times 10^{-2} \div 10^{-1}$, therefore, if we put these numbers into the expression for the theoretical bound in Proposition 2, the latter exceeds 1 after a few timesteps. Furthermore, these values suggest that the near-term devices we have employed are still orders of magnitude away from the strictest quantum fault-tolerant thresholds, at least if we assume, as it may be reasonable, that SPAM errors do not affect the diamond distances by several orders of magnitudes.
 
In addition, we have employed the experimental results of the noise analysis to build a noise model that (classically) simulates the state of the collision model at timestep $n$ by replacing all the CNOT gates in the algorithm with the noisy channels obtained through process tomography. We have found that this noise model predicts the experimental state with a reasonable accuracy (the infidelity between experimental and simulated state is typically lower than 0.1, while the one between experimental and ideal state is between five and ten times higher, see Fig.~\ref{fig:errortimestep1}). Moreover, we have compared it with the built-in noise model provided by IBM that considers also single-qubit gates and qubit relaxation, and discovered that their performance is quite similar. However, our noise model may outperform the IBM one in the presence of very noisy gates (see Fig.~\ref{fig:errortimestep2} in Appendix~\ref{sec:detrimental}), and this might also suggest that the high values of average gate infidelities displayed in Fig.~\ref{fig:erroran2} in Appendix~\ref{sec:detrimental} are more informative than the more optimistic IBM gate errors.


{A crucial issue of open system simulation via interactions with ancillary qubits is the need for a new ancilla at each timestep. In this study, we have employed a train of ancillas that are swapped after each collision, so that a single ancillary qubit is finally interacting with the system qubits (see the discussion in Appendix~\ref{sec:expScheme}). A different solution consists in employing a reset gate to refresh the state of a single ancillary qubit at every timestep. However, our preliminary results based on the reset gate have shown a very quick emergence of decoherence (Fig.~\ref{fig:reset} in Appendix~\ref{sec:further}). Reducing the gate time of the reset gate would therefore be a remarkable improvement for any simulation algorithm making use of ancillas. If this is not possible, enriching the topology of the near-term devices will crucially reduce the number of swap gates necessary to refresh the state of the ancilla for the $n$th collision. Yet another possibility for enhancing the accuracy of the quantum simulation would be working at the pulse level of the near-term devices \cite{Alexander2020}. This would basically correspond to performing an analog quantum simulation instead of a digital one, and has very recently led to improved results on the quantum simulation of many-body unitary systems \cite{stenger2021simulating}.}

To conclude, our experimental outcomes highlight how {near-term quantum computers}, even if far from being ideal, can already give useful and meaningful results on the simulation of {collective} quantum dynamics. Our {findings} are a proof-of-principle demonstration of the potential of the MCM algorithm for the exploration of topical and groundbreaking phenomena such as dissipative quantum phase transitions, quantum synchronization, and dissipative time crystals. Finally, we have demonstrated that gate process tomography,  despite its potential bias due to residual SPAM errors, can give valuable information about the noise features of near-term quantum computers and about the accuracy of the experimentally simulated state. We have shown how this can help us to understand the device limitations, and hence to engineer possible countermeasures.

\acknowledgements{
We would like to thank the anonymous referees of PRX Quantum for giving us valuable suggestions on how to improve the analysis and presentation of our experimental results.
We acknowledge the use of IBM Quantum services for this work {and the Quantum Technologies Platform QTEP (CSIC)}. The views expressed are those of the authors, and do not reflect the official policy or position of IBM or the IBM Quantum team.
M.C., M.A.C.R., G.G.-P., and S.M. acknowledge financial support from the Academy of Finland via the Centre of Excellence program (Project No. 336810 and Project No. 336814). {S.M. and G.G.-P. acknowledge support from the emmy.network foundation under the aegis of the Fondation de Luxembourg. G.G.-P. acknowledges support from the Academy of Finland via the Postdoctoral Researcher program (Project no. 341985).} {
M.C. and R.Z. acknowledge financial support from Centers and Units of Excellence in R\&D (MDM-2017-
0711) and from MICINN/AEI/FEDER and CAIB for projects PID2019-109094GB-C21/AEI/10.13039/501100011033 and PRD2018/47.}
}

\appendix

\section{Distances in quantum information}
\label{sec:distances}
We will introduce here some distance measures for operators on the Hilbert space $\mathcal{H}$ \cite{nielsenchuang,Watrous2018,Watrous2011}.

\begin{definition}[Schatten norms]
The Schatten $p$-norm (with $p\in[1,\infty]$) of an operator $A\in\mathcal{B}(\mathcal{H})$ is defined as:
\begin{equation}
    \normgood{A}_p=\Tr[\left(\sqrt{A^\dagger A}\right)^p]^{\frac{1}{p}}.
\end{equation}
\end{definition}
We will focus on the following two Schatten norms:
\begin{definition}[Trace norm]
The trace norm (or 1-norm) of an operator $A\in\mathcal{B}(\mathcal{H})$ is defined as:
\begin{equation}
\label{eqn:traceNorm}
    \normgood{A}_1=\Tr[\sqrt{A^\dagger A}].
\end{equation}
\end{definition}
\begin{definition}[Operator norm]
The operator norm (or infinity norm) of an operator $A\in\mathcal{B}(\mathcal{H})$ is the standard one in functional analysis, that is:
\begin{equation}
\label{eqn:operatorNorm}
    \normgood{A}_\infty=\max_{\ket{v}\in\mathcal{H}:\,\normgood{v}=1}\normgood{A\ket{v}}.
\end{equation}
\end{definition}
The vector norm (without any subscript) $\normgood{v}$ is the standard Euclidean norm in the Hilbert space $\mathcal{H}$.

Here are some properties of the Schatten norms, in addition to the ones that define a norm, i.e., positivity, being zero only if $A=0$, and fulfilling the triangle inequality $\normgood{A+B}_p\leq \normgood{A}_p+\normgood{B}_p$:
\begin{description}
\item[Unitary invariance] $\normgood{U A V}_p=\normgood{A}_p$ if $U,V$ unitary.
\item[Sub-multiplicativity] $\normgood{AB}_p\leq \normgood{A}_p\normgood{B}_p$.
\item[Monotonicity] $\normgood{A}_1\geq\normgood{A}_p\geq\normgood{A}_q\geq\normgood{A}_\infty$ for $p\leq q$.
\item[H\"{o}lder's inequality] $\normgood{AB}_1\leq \normgood{A}_p\normgood{B}_q$, with $p,q\in[1,\infty]$ (properly generalized) and $1/p+1/q=1$. In particular, $\normgood{AB}_1\leq \normgood{A}_1\normgood{B}_\infty$.
\item[Stability under tensor product] $\normgood{A\otimes\mathbb{I}_{B}}_p=(d_B)^{\frac{1}{p}}\normgood{A}_p$. In particular, $\normgood{A\otimes\mathbb{I}_{B}}_\infty=\normgood{A}_\infty$, $\normgood{A\otimes\mathbb{I}_{B}}_1=d_B\normgood{A}_1$. 
\end{description}

An additional useful figure of merit is the fidelity between two quantum states:
\begin{equation}
\label{eqn:fidelity}
    \mathcal{F}(\rho,\sigma)=\normgood{\sqrt{\rho}\sqrt{\sigma}}_1^2=\left(\Tr[\sqrt{\sqrt{\rho}\sigma\sqrt{\rho}}]\right)^2.
\end{equation}
If $\rho=\ket{\psi}\bra{\psi}$ is pure, then $\mathcal{F}(\rho,\sigma)=\Tr[\rho\sigma]$. The \textit{infidelity} $1-\mathcal{F}(\rho,\sigma)$ is not a well-defined mathematical distance because it does not satisfy the triangle inequality (although it can  be easily turned into a well-defined metric, e.g., by taking its square root \cite{Gilchrist2005}).

\section{Steps of the multipartite collision model}
\label{sec:steps}
We provide here the steps of the multipartite collision model introduced in Ref.~\cite{Cattaneo2021} and discussed in Sec.~\ref{sec:algorithm}. Let us consider the master equation Eq.~\eqref{eqn:gkls} of a multipartite open system whose dynamics we aim to simulate. For simplicity, let us now assume that the subsystems $1,\ldots,M$ are qubits and the Lindblad operators in Eq.~\eqref{eqn:dissipator} are linear combinations of local operators acting on a single subsystem only. That is, $L_k=\sum_{m=1}^{M} F_{m}^{(k)}$, and each $F_{m}^{(k)}$ is local on the $m$th subsystem. Then, the steps of the algorithm can be expressed as follows:
\begin{enumerate}
    \item To each $k=1,\ldots,J$ in Eq.~\eqref{eqn:dissipator} assign an ancillary qubit that will generate the corresponding term of the master equation.
    \item For each $k=1,\ldots,J$, prepare the set of two-qubit quantum gates between the $k$th ancilla and the $m$th subsystem defined by:
    \begin{equation}
    \label{eqn:single_gate}
        U_k^{(m)}(t)=\exp[-i g_It(\lambda_k F_m^{(k)}\sigma_k^++h.c.)],
    \end{equation}
    where $\lambda_k$ is a dimensionless parameter that is defined by $\Gamma_k=\abs{\lambda_k}^2$, while the gate time $t$ will eventually assume two values only, namely either $\Delta t$ or $\Delta t/2$, where $\Delta t$ is the timestep of the algorithm. $g_I$ is a coupling constant which defines the collision strength, and in this paper it is fixed to $g_I=\Delta t^{-1/2}$ \cite{Cattaneo2021}.
    \item Compose these quantum gates into a single unitary evolution $U_k(\Delta t)$ as follows:
    \begin{equation}
    \label{eqn:suzuki-Trotter}
        U_k(\Delta t)=\prod_{m=1}^M U_k^{(M-m+1)}(\Delta t/2)\prod_{m'=1}^M U_k^{(m')}(\Delta t/2).
    \end{equation}
    Note that the two-qubit gate $U_k^{(1)}(\Delta t)$ can be implemented as a single gate lasting for $\Delta t$, while all the other gates last for $\Delta t/2$ and are executed twice during a single timestep. Eq.~\eqref{eqn:suzuki-Trotter} expresses the second-order Suzuki-Trotter decomposition of the interaction between the subsystems and the ancilla.
    \item Follow the same procedure for each ancilla $k=1,\ldots,J$, and insert each gate sequence in a total unitary operator, where their order of execution does not matter:
    \begin{equation}
    \label{eqn:evolution_int_complete}
        U_I(\Delta t)=\prod_{k=1}^{d^{2M}-1}U_k(\Delta t).
    \end{equation}
    \item Introduce a sequence of gates that simulate the effective Hamiltonian $H_S$, which is a free system Hamiltonian, during the timestep $\Delta t$:
    \begin{equation}
    \label{eqn:evolution_sim}
    U_{sim}(\Delta t)=U_S(\Delta t)\circ U_I(\Delta t),
    \end{equation}
    with $U_S(\Delta t)=\exp[-i H_S\Delta t].$ Note that the latter step amounts to simulating the closed-system dynamics driven by the Hamiltonian $H_S$, which is a well-known task in quantum computing since the seminal paper by Lloyd \cite{Lloyd1996}. 
    \item Initialize the ancillary qubits in the ground state expressed by $\rho_E=\bigotimes_{k=1}^{J}\ket{0}_k\!\bra{0}$. Initialize the system qubits in the initial state of the open dynamics $\rho_S(0)$, as introduced in Eq.~\eqref{eqn:evolution}.
    \item Implement a single timestep $\Delta t$ of the algorithm by making the system and ancillary qubits evolve through the sequence of gates contained in the operator $U_{sim}(\Delta t)$. Then, to obtain the information on the state of the system only, trace out the degrees of freedom of the ancillas (the environment):
    \begin{equation}
    \label{eqn:quantumMap}
    \phi_{\Delta  t}[\rho_S]=\Tr_E[U_{sim}(\Delta t)\rho_S\otimes\rho_E U_{sim}^\dagger(\Delta t)].
    \end{equation}
    $\phi_{\Delta t}$ is the quantum map associated with a single timestep of the multipartite collision model.
    \item To implement $n$ timesteps of the algorithm, apply the quantum map $\phi_{\Delta t}$ $n$ times. That is, repeat the sequence of gates contained in $U_{sim}(\Delta t)$ for $n$ times using the same system qubits and a new set of fresh ancillas, initialized in the ground state $\rho_E$, for each timestep. The simulation of the dynamics until time $t$ requires $n=t/\Delta t$ repetitions, where the timestep of the algorithm $\Delta t$ should be chosen as small as the experimental conditions allow for.
\end{enumerate}

\section{Analytical solution of the dynamics}
\label{sec:anSol}
Here, we provide the analytical solution of the collective dynamics for the different initial states we consider in this work. Namely, we will obtain $\rho_S(t)=\exp{\mathcal{L}t}[\rho_S(0)]$, where $\mathcal{L}$ is given by Eq.~\eqref{eqn:LiouvillianSup}, and for $\rho_S(0)=\rho_{\mathrm{sup}},\,\rho_{\mathrm{sub}}$ and $\rho_{ee}=\ket{ee}\bra{ee}$, which have been introduced in Sec.~\ref{sec:subSup}. We will also find the evolution of $\rho_{\mathrm{sub}}$ when the dissipator is local and incoherent, that is, in the absence of the decoherence-free subspace. 

A straightforward calculation yields $\mathcal{L}[\rho_{\mathrm{sub}}]=0$, therefore 
\begin{equation}
    \label{eqn:evSub}
    \exp{\mathcal{L}t}[\rho_{\mathrm{sub}}]=\rho_{\mathrm{sub}},
\end{equation} i.e., $\rho_{\mathrm{sub}}$ is a steady state of the dynamics and lives in a decoherence-free subspace.

As for the superradiant state, we observe:
\begin{equation}
\label{eqn:eqSup}
    \mathcal{L}[\rho_{\mathrm{sup}}]=-2\gamma (\rho_{\mathrm{sup}}-\rho_{gg}),\quad \mathcal{L}[\rho_{gg}]=0,
\end{equation}
where $\rho_{gg}=\ket{gg}\bra{gg}$ is the ground state, which is stationary because the master equation~\eqref{eqn:LiouvillianSup} is at zero temperature. Therefore, the operator $\rho_{\mathrm{sup}}-\rho_{gg}$ is an eigenvector of the Liouvillian with eigenvalue $-2\gamma$. We finally obtain:
\begin{equation}
\begin{split}
    \label{eqn:evSup}
    \exp{\mathcal{L}t}[\rho_{\mathrm{sup}}]&=\exp{\mathcal{L}t}[\rho_{\mathrm{sup}}-\rho_{gg}]+\rho_{gg}\\
    &=e^{-2\gamma t}\rho_{\mathrm{sup}}+(1-e^{-2\gamma t})\rho_{gg}.
\end{split}
\end{equation}
Using the above result, we immediately find the formula for the intensity of the superradiant emission in Eq.~\eqref{eqn:intensity}.

Computing the evolution of $\rho_{ee}$ is slightly more involved. First of all, we calculate:
\begin{equation}
    \mathcal{L}[\rho_{ee}]=2\gamma(\rho_{\mathrm{sup}}-\rho_{ee}).
\end{equation}
Then, we can merge the results of the above equation and Eq.~\eqref{eqn:eqSup} into a single system of linear differential equations written as:
\begin{equation}
    \frac{d}{dt}\mathbf{v}(t)=M \mathbf{v}(t), 
\end{equation}
where $ \mathbf{v}(t)=(\exp\mathcal{L}t[\rho_{gg}],\exp\mathcal{L}t[\rho_{\mathrm{sup}}],\exp\mathcal{L}t[\rho_{ee}])^T$, 
\begin{equation}
    M=\gamma\begin{pmatrix}
    0 & 0 & 0 \\
    2 & -2 & 0 \\
    0 & 2 & -2 
    \end{pmatrix}.
\end{equation}
One way to tackle this problem is to solve each differential equation starting from the trivial one for the steady state, and then to insert this solution into the following equation, which will now be independent and affine, and so on. Another instructive (and more general) way to find the solution of the dynamics is to compute $\mathbf{v}(t)=\exp (Mt)\mathbf{v}(0)$ by getting the Jordan–Chevalley decomposition of $M$ \cite{lang2002}. Indeed, $M$ has the eigenvalue 0 with eigenvector $(1,1,1)^T$, and the eigenvalue $-2$ with multiplicity $2$ and a one-dimensional eigenspace spanned by $(0,0,1)^T$. That is, $M$ is not diagonalizable. So, we need to obtain its Jordan decomposition as $M= P J_M P^{-1}$, with:
\begin{equation}
        J_M= \gamma\begin{pmatrix}
        0 & 0 & 0 \\
        0 & -2 & 1\\
        0 & 0 & -2 \\
        \end{pmatrix},\quad
        P= \begin{pmatrix}
        1 & 0 & 0 \\
        1 & 0 & 1\\
        1 & 2 & 0 \\
        \end{pmatrix}.
\end{equation}
Then, the Jordan-Chevalley decomposition is trivially expressed by $J_M=D_M+N_M$, where $D_M$ is the diagonal matrix with the same elements of $J_M$ on the diagonal, while $N_M$ is a matrix whose only non-zero element is the off-diagonal $1$ in $J_M$. $N_M$ is nilpotent ($N_M^2=0$), and $[N_M,D_M]=0$. Then, after some simple matrix algebra we find 
\begin{equation}
\begin{split}
        e^{Mt}&=e^{P J_M P^{-1} t}=P e^{D_Mt} e^{N_Mt} P^{-1}\\
        & = \begin{pmatrix}
        1 & 0 & 0 \\
         1-e^{-2\gamma t} & e^{-2\gamma t} & 0 \\
          1- e^{-2\gamma t}(1+2\gamma t) &  2\gamma t e^{-2\gamma t} &   e^{-2\gamma t}
        \end{pmatrix}.
\end{split}
\end{equation}

Finally, we obtain the evolution of $\rho_{ee}$:
\begin{equation}
\begin{split}
    \label{eqn:evEE}
    \exp{\mathcal{L}t}[\rho_{ee}]=&e^{-2\gamma t}\rho_{ee} + 2\gamma t e^{-2\gamma t}\rho_{\mathrm{sup}}\\
    &+(1-e^{-2\gamma t}(1+2\gamma t))\rho_{gg}.
\end{split}
\end{equation}
Once again, the enhanced decay rate $2\gamma$ is a signature of collective effects in the dynamics.

Let us now compute the local incoherent evolution of $\rho_{\mathrm{sub}}$. The Liouvillian driving this type of dynamics is:
\begin{equation}
\label{eqn:liouvLoc}
    \mathcal{L}_{loc}[\rho_S]=\sum_{j=1,2}\gamma\left(\sigma_j^-\rho_S\sigma_j^+-\frac{1}{2}\{\sigma_j^+\sigma_j^-,\rho_S\}\right),
\end{equation}
where, for simplicity, we have taken the same decay rate $\gamma$ for both qubits. Straightforwardly, $\mathcal{L}_{loc}[\rho_{\mathrm{sub}}]=\gamma( \rho_{gg}-\rho_{\mathrm{sub}})$, therefore
\begin{equation}
\label{eqn:evLoc}
    \exp\mathcal{L}_{loc}t[\rho_{\mathrm{sub}}]=e^{-\gamma t}\rho_{\mathrm{sub}}+(1-e^{-\gamma t})\rho_{gg}.
\end{equation}
As expected, the subradiant state decays toward the ground with the standard incoherent rate $\gamma$.

\section{Proofs of the results in Sec.~\ref{sec:errorBoundTheo}}

\label{sec:proofs}

\subsection{Diamond distance $\epsilon_s^\Diamond$ for the ideal case}
We want to prove the bound
\begin{equation}
    \normgood{\phi_{\Delta t}-\exp \mathcal{L}\Delta t}_\Diamond\leq \mathcal{B}_{1\rightarrow 1}.
\end{equation}
To check its validity, note that the bounds we need to estimate are, according to the Supplemental Material (SM) of Ref.~\cite{Cattaneo2021},
\begin{equation}
    \normgood{\mathcal{L}}_\Diamond,\qquad \normgood{\mathsf{R}_c\left(\Tr_E[U(\Delta t)\cdot\otimes\rho_E U^\dagger(\Delta t)]\right)}_\Diamond,
\end{equation}
where $\mathsf{R}_c$ is a remainder of the expansion of the MCM unitaries (see the original paper \cite{Cattaneo2021} for details). For our purposes, it is sufficient to assume that it is a polynomial function of operators. 

Let us first show that the bound on the $1\rightarrow 1$ norm of the Liouvillian found is valid also for the diamond norm. This result is also stated in Ref.~\cite{Werner2016a} (SM). The crucial point is that, despite the $1\rightarrow 1$ having the undesirable property of not scaling well in the presence of a tensor product, by means of the H\"{o}lder's inequality introduced in Sec.~\ref{sec:FiguresOfMerit}, the error bound can be written as a function of infinity norms only, which are stable. Therefore, estimating the same bound with the diamond norm leads to the same equations. Let us show this.

\begin{widetext}
In the following, $\mar{\mathcal{I}}_A$ will be the identity superoperator on a copy of the space of the bounded operators on the Hilbert space of the system $\mathcal{B}(\mathcal{H})$, according to the definition of the diamond norm in Eq.~\eqref{eqn:diamondNorm}. We evaluate:
\begin{equation}    
\label{eqn:boundLiouvillianDiamond}
    \begin{split} \normgood{\mathcal{L}\otimes\mar{\mathcal{I}}_A[\rho]}_1&=-i\normgood{[H\otimes\mathbb{I}_A,\rho]}_1+\sum_k\left(\normgood{L_k\otimes\mathbb{I}_A \rho L_k^\dagger\otimes\mathbb{I}_A-\frac{1}{2}\{L_k^\dagger L_k\otimes\mathbb{I}_A\}}_1 \right)\\
    &\leq  2 (\normgood{H}_\infty+\sum_k \normgood{L_k}_\infty^2),
    \end{split}
\end{equation}
under the assumption that $\normgood{\rho}_1=1$. We have used the H\"{o}lder's inequality and the multiplicativity of the infinity norm. The above bound is the same as for the $1\rightarrow 1$ norm. Indeed, evaluating everything on a basis of the extended Hilbert space we observe:
\begin{equation}
\mathcal{L}\otimes\mar{\mathcal{I}}_A[\ket{\psi_j\psi_k}\bra{\psi_l\psi_m}]=\mathcal{L}[\ket{\psi_j}\bra{\psi_l}]\otimes\ket{\psi_k}\bra{\psi_m},
\end{equation}
and we can write $\rho$ as
\begin{equation}
\rho=\sum_{j,k,l,m}a_{jklm}\ket{\psi_j\psi_k}\bra{\psi_l\psi_m}.
\end{equation}
Therefore,
\begin{equation}
\begin{split}
\mathcal{L}\otimes\mar{\mathcal{I}}_A[\rho]=& \sum_{j,k,l,m}a_{jklm} \left(-i[H,\ket{\psi_j}\bra{\psi_l}]\otimes\ket{\psi_k}\bra{\psi_m}+\sum_k(L_k\ket{\psi_j}\bra{\psi_l} L_k^\dagger-\ldots)\otimes\ket{\psi_k}\bra{\psi_m}\right),
\end{split}
\end{equation}
showing that Eq.~\eqref{eqn:boundLiouvillianDiamond} is correct. This is basically due to the linearity of quantum channels.

Equivalently, for the remainder:
\begin{equation}
    \begin{split}
&\sum_{j,k,l,m}a_{jklm} \Tr_E[U(\Delta t)\ket{\psi_j}\bra{\psi_l}\otimes\rho_E U^\dagger(\Delta t)]\otimes\ket{\psi_k}\bra{\psi_m}\\
=&\sum_{j,k,l,m}a_{jklm} \Tr_E[U(\Delta t)\otimes\mathbb{I}_A\ket{\psi_j\psi_k}\bra{\psi_l\psi_m}\otimes\rho_E U^\dagger(\Delta t)\otimes\mathbb{I}_A]
=\Tr_E[U(\Delta t)\otimes\mathbb{I}_A \rho \otimes\rho_E U^\dagger(\Delta t)\otimes\mathbb{I}_A].
    \end{split}
\end{equation}
Therefore, when evaluating the trace norm of the above expression, we can still find a bound that only depends on $\normgood{U(\Delta t)\otimes\mathbb{I}_A}_\infty$, that is, on $\normgood{U(\Delta t)}_\infty$. This is the same bound found in Ref.~\cite{Cattaneo2021}, expressed in Eq.~\eqref{eqn:bound} of the main text.

\subsection{Diamond distance $\epsilon_m^*$ between ideal and noisy MCM map}
The estimation of an upper bound for $\epsilon_m^*$ can be performed as follows:
\begin{equation}
\label{eqn:boundMCMnoisyComputing}
\begin{split}
    \normgood{\phi_{\Delta t}^*-\phi_{\Delta t}}_\Diamond=&\normgood{\Tr_E\left[\prod_j \mathcal{E}_j\prod_i\mar{\mathcal{I}}_S\otimes\mathcal{G}_i[\cdot\otimes\rho_E]\right]-\Tr_E\left[\prod_j \mathcal{U}_j[\cdot\otimes\rho_E]\right]}_\Diamond\\
    =&\normgood{\Tr_E\left[\left(\prod_j \mathcal{E}_j\prod_i\mar{\mathcal{I}}_S\otimes\mathcal{G}_i-\prod_j \mathcal{U}_j\prod_i\mar{\mathcal{I}}_{SE}\right)[\cdot\otimes\rho_E]\right]}_\Diamond\\
    =&\sup_{\rho_{SA}:\normgood{\rho_{SA}}_1=1}\normgood{\Tr_E\left[\left(\prod_j \mathcal{E}_j\prod_i\mar{\mathcal{I}}_S\otimes\mathcal{G}_i-\prod_j \mathcal{U}_j\prod_i\mar{\mathcal{I}}_{SE}\right)\otimes\mar{\mathcal{I}}_A[\rho_{SA}\otimes\rho_E]\right]}_1\\
    \leq&\sup_{\rho_{SA}:\normgood{\rho_{SA}}_1=1}\normgood{\left(\prod_j \mathcal{E}_j\prod_i\mar{\mathcal{I}}_S\otimes\mathcal{G}_i-\prod_j \mathcal{U}_j\prod_i\mar{\mathcal{I}}_{SE}\right)\otimes\mar{\mathcal{I}}_A[\rho_{SA}\otimes\rho_E]}_1\\
    \leq&\sup_{\rho_{SA}\otimes\rho_E:\normgood{\rho_{SA}}_1=1,\normgood{\rho}_E=1}\normgood{\left(\prod_j \mathcal{E}_j\prod_i\mar{\mathcal{I}}_S\otimes\mathcal{G}_i-\prod_j \mathcal{U}_j\prod_i\mar{\mathcal{I}}_{SE}\right)\otimes\mar{\mathcal{I}}_A[\rho_{SA}\otimes\rho_E]}_1\\
    \leq&\sup_{\rho_{SAE}:\normgood{\rho_{SAE}}_1=1}\normgood{\left(\prod_j \mathcal{E}_j\prod_i\mar{\mathcal{I}}_S\otimes\mathcal{G}_i-\prod_j \mathcal{U}_j\prod_i\mar{\mathcal{I}}_{SE}\right)\otimes\mar{\mathcal{I}}_A[\rho_{SAE}]}_1\\
    \leq&\sup_{\rho_{SAEB}:\normgood{\rho_{SAEB}}_1=1}\normgood{\left(\prod_j \mathcal{E}_j\prod_i\mar{\mathcal{I}}_S\otimes\mathcal{G}_i-\prod_j \mathcal{U}_j\prod_i\mar{\mathcal{I}}_{SE}\right)\otimes\mar{\mathcal{I}}_A\otimes\mar{\mathcal{I}}_B[\rho_{SAEB}]}_1\\
    =&\normgood{\prod_j \mathcal{E}_j\prod_i\mar{\mathcal{I}}_S\otimes\mathcal{G}_i-\prod_j \mathcal{U}_j\prod_i\mar{\mathcal{I}}_{SE}}_\Diamond
    \leq\sum_j\normgood{\mathcal{E}_j-\mathcal{U}_j}_\Diamond+\sum_i\normgood{\mathcal{G}_i-\mar{\mathcal{I}}_{E}}_\Diamond.
\end{split}
\end{equation}
\end{widetext}
Note that we are employing the subscript ``$A$'' to denote states and operators that belong or act on a copy of the Hilbert space of the system, while ``$B$'' denotes states and operators that belong or act on a copy of the Hilbert space of the ancillary qubits of the collision model, according to the definition of the diamond norm in Eq.~\eqref{eqn:diamondNorm}. In contrast, the subscripts ``$S$'' and ``$E$'' indicate respectively the ``original'' Hilbert space of the system and the Hilbert space of the ancillary qubits. To obtain the final result in Eq.~\eqref{eqn:boundMCMnoisyComputing}, we have employed the fact that $\normgood{\Tr_E[O]}_1\leq \normgood{O}_1$ \cite{Lidar2008}, the sub-multiplicativity of the diamond norm, as in Eq.~\eqref{eqn:submultiplicativity}, and the fact that the diamond norm defined in Eq.~\eqref{eqn:diamondNorm} with the tensor product of the identity operator over a copy of the Hilbert space (in our case of system + ancillary qubits of the MCM) is the maximal one \cite{Watrous2018}.

{Finally, note that the diamond distance between the map $\mathcal{G}_i$ and the identity does not assume that the ancilla is ideally initialized in the ground state. So, Eq.~\eqref{eqn:boundMCMnoisyComputing} is valid also for scenarios with different choices of initial states. Moreover, if the quantum channel describing the noise for the state preparation on the backend is not known, one may tighten the bound by minimizing $\normgood{\mathcal{G}_i-\mar{\mathcal{I}}_{E}}_\Diamond$ over all the channels $\mathcal{G}_i$ that map the ideal initial state of the ancilla into the observed noisy state.}

\section{Experimental scheme and methods}
\label{sec:expScheme}

\subsection{Topology of the backend}
The IBM quantum computers are often recalibrated to optimize their performance. That is, their qubit and gate parameters are often modified at the level of the hardware. As a consequence, the gate and readout errors change after every calibration. For our purposes, we want to run the above-listed protocols with a fixed set of experimental parameters. So, we need to employ a near-term device with a sufficient time interval between two calibration procedures (running all the necessary protocols listed in Sec.~\ref{sec:IBM} takes roughly 3 hours). The 16-qubits backend \texttt{ibmq\_guadalupe} is very stable in this sense (it is usually re-calibrated once per day), and so we have chosen it for running our algorithm.

\begin{figure}[t!]
    \centering
    \includegraphics[scale=0.48]{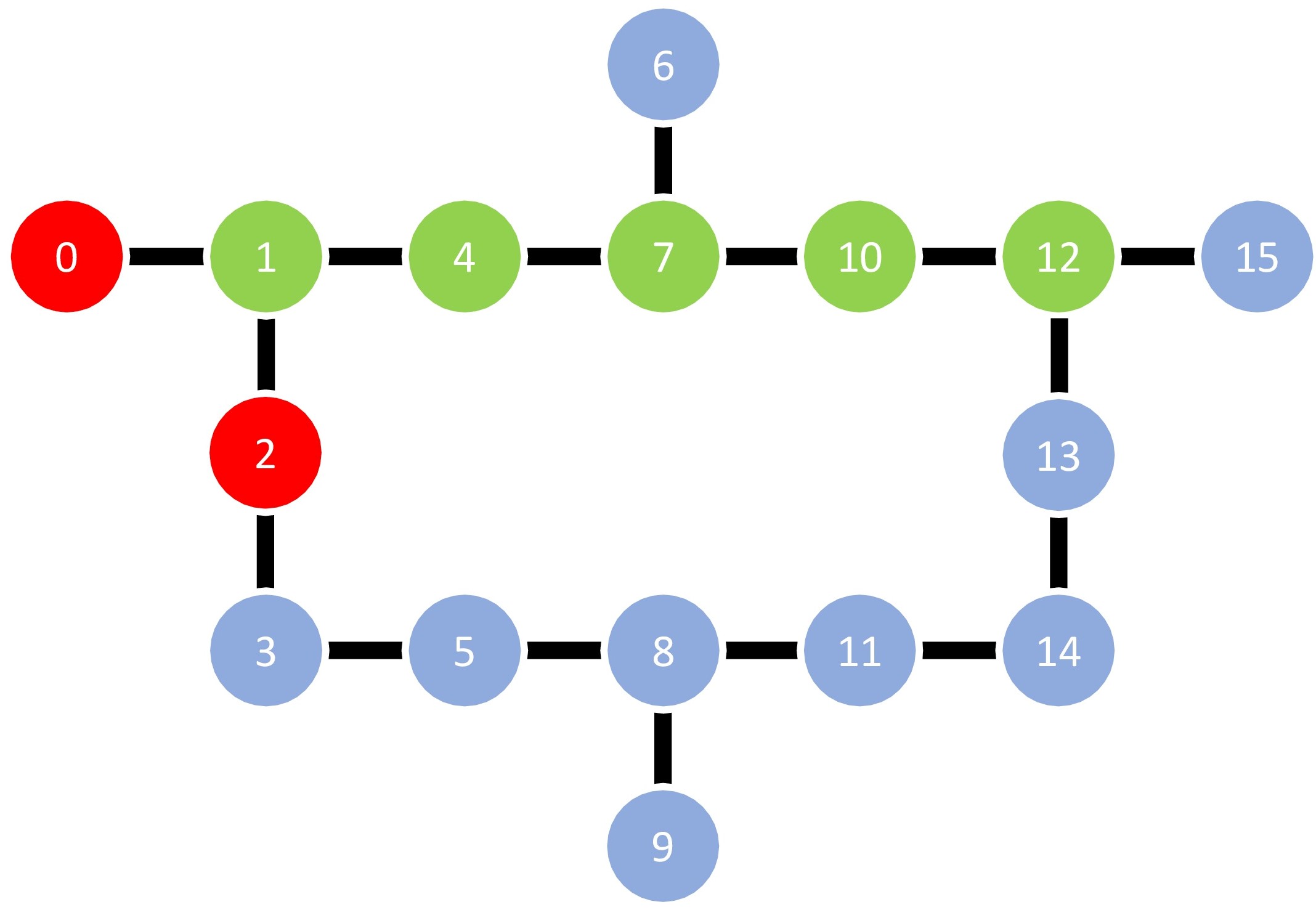}
    \caption{Topology of \texttt{ibmq\_guadalupe} with 16 qubits. The links denote the possibility of implementing a direct two-qubit gate between a pair of qubits. The red (green) qubits are employed as system (ancillary) qubits in the simulation of dissipative collective effects.}
    \label{fig:guadalupe}
\end{figure}

When we write the (Qiskit) code to run a quantum algorithm on a near-term device, we must always take into account the topology of the latter, because it constraints the operations we can actually perform on the platform: only CNOTs between nearby qubits can be directly implemented. The topology of  \texttt{ibmq\_guadalupe} is depicted in Fig.~\ref{fig:guadalupe}.

For the experiments presented in Sec.~\ref{sec:IBM}, we have used the following sets of qubits on \texttt{ibmq\_guadalupe}: the system qubits are $(0,2)$. The ancillary qubits for the collective dynamics are $(1,4,7,10,12)$. We have employed the same set of ancillary qubits for the local dissipative dynamics on the system qubit $0$, while the ancillas for the local decay of the system qubit $2$ are $(3,5,8,11,14)$.

Note that the ancillary qubit 1 (and 3 for the local decay of the system qubit 2) is the only one that is directly linked to both the system qubits. So, we are only allowed to implement the CNOTs 0--1, 1--0, 1--2 and 2--1 to evolve the state of the system. To make the system qubits interact with the remaining ancillas, we need to swap the states of the ancillary qubit 1 with the one of the ancillary qubit 4, then 4 with 7 and again 1 with 4, then 7 with 10, and so on. This requires an additional number of CNOT gates (three per each swap) that increases at each timestep. Indeed, given a train of ancillary qubits from $1$ (which is directly linked to the system qubits) to $n$, we need exactly $n-1$ swaps to make the $n$-th ancilla interact with the system. This is a well-known issue of near-term devices with limited connectivity \cite{Garcia-Perez2020}. In the next subsection, we will show how we have been able to optimize the number of necessary gates by employing, under certain circumstances, two CNOTs only to perform a single swap. 

Possible ways to tackle this problem in the near future may be: i) Adding more qubit-qubit links to the topology of the backend, so as to reduce the number of swaps necessary to connect a distant ancilla with the system qubits; ii) Implementing a fast and efficient reset gate to be applied on the closest ancillary qubit (qubit 1 in Fig.~\ref{fig:guadalupe}) at every timestep. In fact, by employing a reset gate, we would need a single ancillary qubit only to be re-initialized on the ground state after each collision, or equivalently, thinking of more complex physical problems, one ancillary qubit for each Lindblad operator in the dissipator in Eq.~\eqref{eqn:dissipator} \cite{Cattaneo2021}, according to the discussion in Sec.~\ref{sec:multipartiteCM}. The reset gate is already available on \texttt{ibmq\_guadalupe}. However, it is a very slow gate and decoherence rapidly emerges after a couple of applications thereof, as we will show in Appendix~\ref{sec:further}. Therefore, improvements on the reset gate time would be extremely beneficial for the quantum simulation of open systems via MCM.

\subsection{Algorithm implementation and optimization}

\begin{figure*}
    \centering
    \subfloat[]{%
    \includegraphics[scale=0.55]{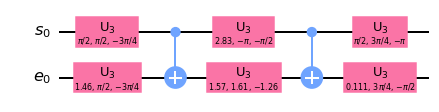}}
    \subfloat[]{%
    \includegraphics[scale=0.45]{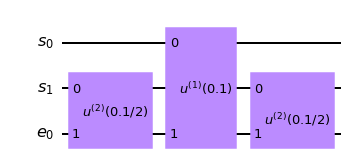}}\\
    \subfloat[]{%
    \includegraphics[scale=0.45]{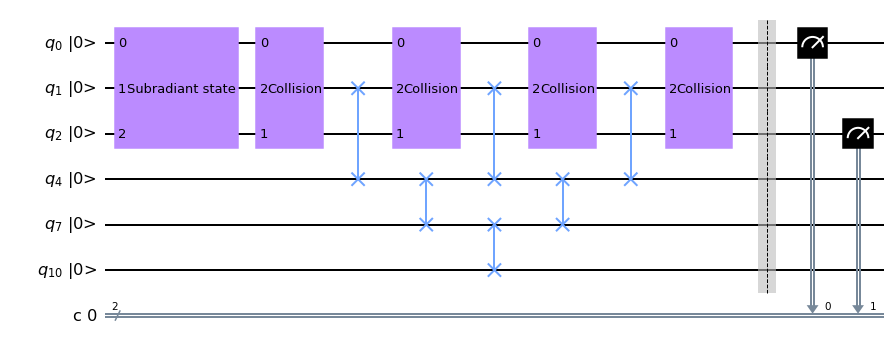}
    }
    \caption{(a): Decomposition of the gate $U^{(1)}(0.1)$ defined in Eq.~\eqref{eqn:gateSupSub} on the IBM quantum computer, applied on a system qubit $s_0$ and an ancillary qubit $e_0$. The gates $U_3$ are single-qubit rotations defined by the three Euler angles given in the figure, and are elementary gates available on the IBM quantum computers \cite{Qiskit}. Two CNOT gates are also necessary to implement the system-ancilla interaction. (b): Scheme of a single collision of the MCM with $\Delta t=0.1$ on the IBM quantum computer. The gates $U^{(1)}(0.1)$ and $U^{(2)}(0.1/2)$ are defined in Eq.~\eqref{eqn:gateSupSub}, and their decomposition into elementary gates is given in subfigure~(a). (c): Experimental circuit scheme to implement four collisions of the MCM on \texttt{ibmq\_guadalupe}, starting from the subradiant state. All the qubits are initialized in the ground state, and the system qubits are then prepared in the subradiant state using the environment qubit 1 as an ancilla (there is no direct interaction between them). After each collision, as given by the decomposition in subfigure~(b), a new fresh state of the ancillary qubit 1 is prepared by swapping the latter with the remaining ancillas of the train.}
    \label{fig:algorithm}
\end{figure*}

The circuit scheme to implement the algorithm discussed in Sec.~\ref{sec:subSup} on \texttt{ibmq\_guadalupe} is shown in Fig.~\ref{fig:algorithm}. The unitary interaction between a single system qubit and the ancilla, i.e., Eq.~\eqref{eqn:single_gate} of the MCM, or equivalently $U^{(i)}(\Delta t)$ in Eq.~\eqref{eqn:gateSupSub} for the simulation of super- and sub-radiance, is depicted in Fig.~\ref{fig:algorithm}(a). To implement it on the hardware, two CNOT gates and some single-qubit rotations are required. Since the gate error of single-qubit rotations is one or two orders of magnitude lower than the CNOT error \cite{Qiskit} (typically, the infidelity of 1-qubit gates is of the order of $10^{-4}$), the CNOTs in the circuit bring the largest contribution to the overall error of the algorithm.

\begin{figure*}
\centering
  \includegraphics[width=.75\textwidth]{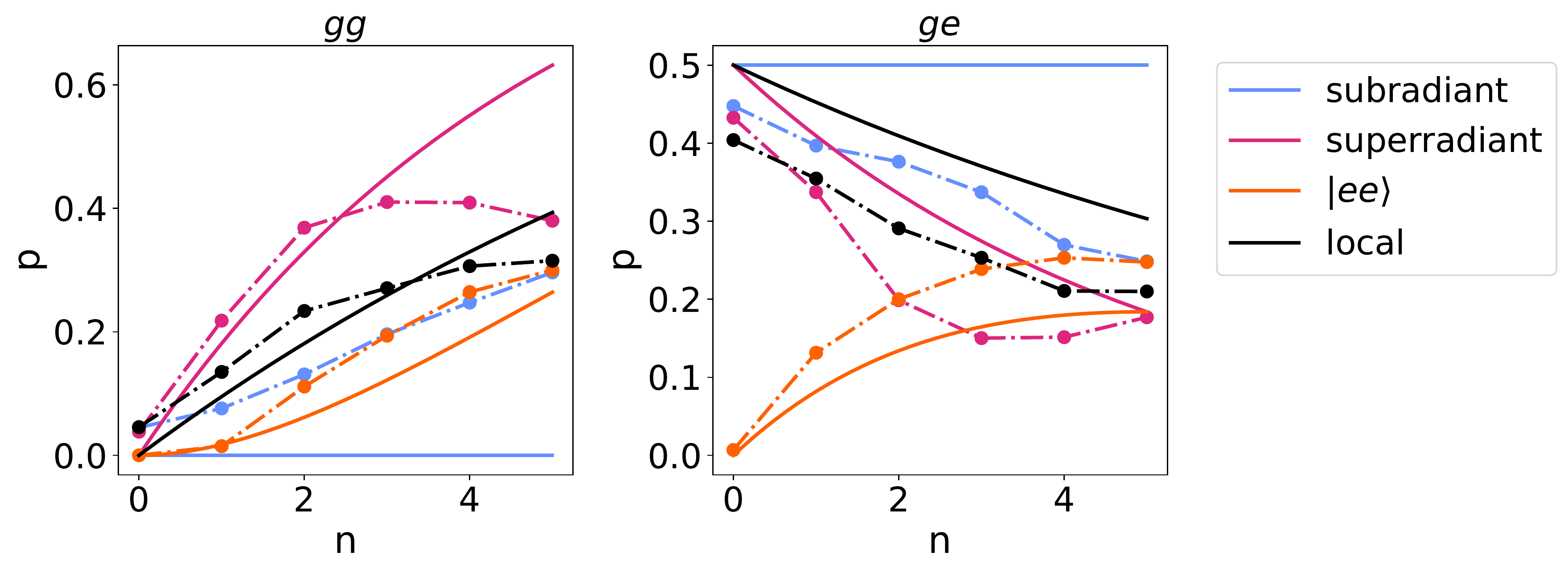}\\
  \includegraphics[width=.75\textwidth]{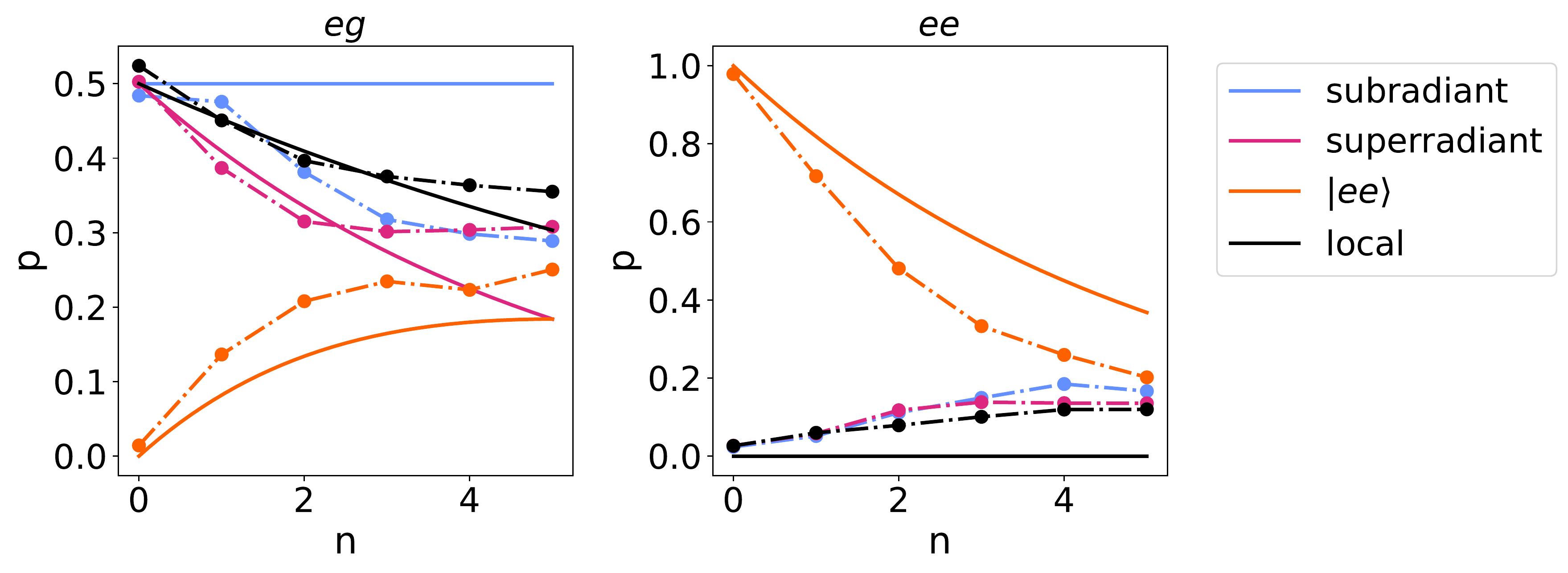}
    \caption{Probability of finding the states $\ket{gg}$, $\ket{ge}$, $\ket{eg}$, $\ket{ee}$ through a projective measurement in the {computational} basis as a function of the number of steps in the MCM  (second set of results), when the initial state is the subradiant state (light blue), superradiant state (magenta), $\ket{ee}$ (orange), and when the dynamics is local starting from the subradiant state (black). Solid lines: theoretical prediction based on the master equation. Markers of the dash-dotted lines: experimental values. The results have been obtained as averages over 37 realizations of the protocol, and the error bars are within the markers.}
    \label{fig:countsTwo}
   \end{figure*}
   
Fig.~\ref{fig:algorithm}(b) represents a single collision between the system qubits and the ancilla, which is composed of the three applications of the gates described in Fig.~\ref{fig:algorithm}(a), according to the second-order Suzuki-Trotter decomposition of the MCM in Eq.~\eqref{eqn:suzuki-Trotter}. Finally, Fig.~\ref{fig:algorithm}(c) shows the implementation of the algorithm on \texttt{ibmq\_guadalupe}, starting from the subradiant state and up to $n=4$ collisions. The preparation of the subradiant and superradiant states requires one CNOT and some single-qubit gates \cite{nielsenchuang,Qiskit}. In addition, due to the topology of the backend in Fig.~\ref{fig:guadalupe}, there is no direct link between the system qubits 0 and 2. Therefore, we need to employ the ancillary qubit 1 to prepare a Bell state between qubits 0 and 1, and then to swap the state of the qubit 1 with the one of qubit 2. In contrast, the preparation of $\ket{ee}$ only needs two local $X$ gates, and is therefore way less noisy.  

As shown in Fig.~\ref{fig:algorithm}(c), after the state preparation we are ready to implement a single timestep of the MCM by applying the routine in Fig.~\ref{fig:algorithm}(b) to the system qubits and the ancilla. To simulate further collisions, we need to re-initialize the state of the ancillary qubit 1 by swapping it with the fresh ancillas in the train (4,7,10,12). We finally measure the state of the system qubits to reconstruct the statistics of the outcomes after the $n$th collision.

The state preparation, the interactions for every timestep and the swaps in the train of ancillas require a large number of noisy CNOT gates. However, we can optimize this number by noticing that either the first or the last CNOT of the usual three-CNOT constructed swap gate
is the identity gate when one of the qubits is in the ground state: if $C_{01}$ is the CNOT with qubit 0 as control and qubit 1 as target, while $S_{01}$ is the swap gate between these qubits, then $S_{01}\ket{0}\otimes\ket{\psi}=C_{01}C_{10}\ket{0}\otimes\ket{\psi}$. That is, we need two CNOTs only to implement the swap gate if one of the qubits starts in the ground state. {Ideally, all the swap gates in the algorithm act at least on one qubit in the ground state. However, we have verified that the 2-CNOT swap is reliable only when the qubit in $\ket{0}$ is a fresh ancilla that has not been manipulated before. Otherwise, the error coming from the fact that this qubit is not exactly in $\ket{0}$ (due to noise and imperfections in the algorithm implementation) jeopardizes the advantage of using one CNOT less. As a consequence, we can remove only 5 CNOT gates from the actual protocol (one for each ancillary qubit).} Finally, note that employing a reset gate would drastically reduce the number of two-qubit gates in the algorithm. Indeed, the train of ancillas at the $n$th collision requires $(n-1)\times(3n-1)$ CNOTs. 

\subsection{Methods to estimate the figures of merit}
The outcome probabilities displayed in Figs.~\ref{fig:countsOne} and~\ref{fig:countsTwo} have been obtained through standard projective measurements in the {computational} basis, which are available on the IBM quantum computer, as in the circuit scheme of Fig.~\ref{fig:algorithm}(c). The results have been computed as averages over 37 realizations of the protocol, and each realization had 8192 shots (i.e., repetitions of the algorithm). We have found that the standard deviation over the 37 realizations always leads to small error bars that are within the markers shown in the plots. 

    \begin{figure*}
  \includegraphics[scale=0.32]{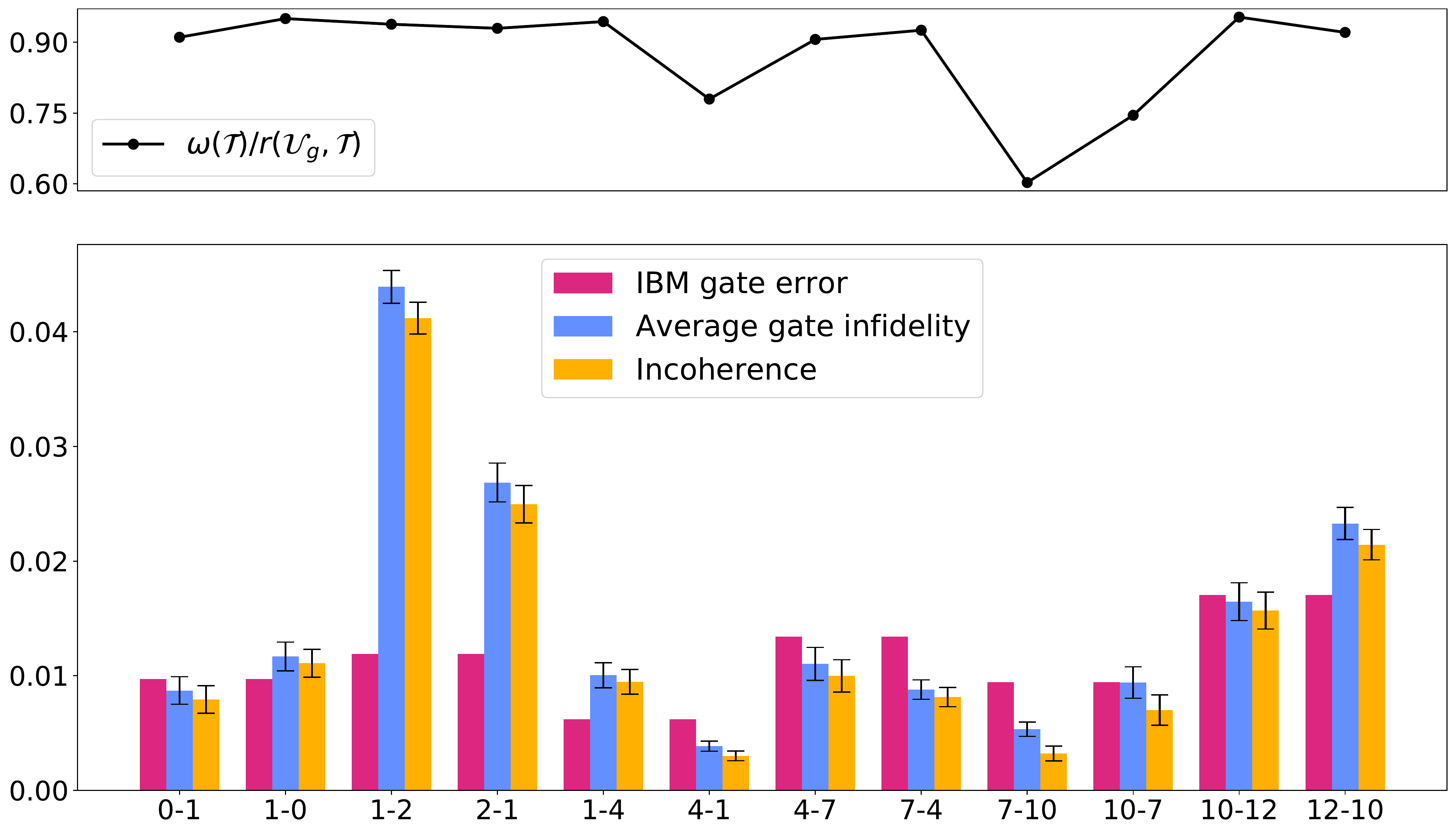}%
   \caption{Error analysis of the set of CNOT gates in \texttt{ibmq\_guadalupe} employed for the simulation of the  second set of results in Fig.~\ref{fig:countsTwo}. The tick 0--1 on the x axis corresponds to the CNOT gate where ``qubit 0'' in \texttt{ibmq\_guadalupe} is the control qubit and ``qubit 1'' is the target. Lower plot: gate error provide by IBM (magenta), experimental average gate infidelity $r(\mathcal{U}_g,\mathcal{T})$ via full process tomography (light blue), and incoherence $\omega(\mathcal{T})$ (yellow), as defined in Eq.~\eqref{eqn:incoherence}. The error bars on the values of $r(\mathcal{U}_g,\mathcal{T})$ and $\omega(\mathcal{T})$ are the standard deviations of 100 realizations of a random sample of the experimental data via bootstrapping. Upper plot: ratio between the incoherence and the experimental average gate infidelity of each CNOT gate.}
    \label{fig:erroran2}
\end{figure*}

The state tomography after each collision (Figs.~\ref{fig:errortimestep1} and~\ref{fig:errortimestep2}) and the process tomography of the CNOT gates (Figs.~\ref{fig:erroran1} and~\ref{fig:erroran2}) have been computed by running the proper tomographic circuits on the backend, which are available on \texttt{qiskit.ignis} \cite{Qiskit}. Then, the results have been fitted to reconstruct either the density matrix of the system through the Qiskit class \texttt{StateTomographyFitter}, or the Choi matrix of the process through the Qiskit class \texttt{ProcessTomographyFitter}. The results have then been resampled 100 times via bootstrapping, after which we have computed the standard deviations we used for the error bars.

The trace distance and the fidelity between ideal and simulated states at each timestep have been computed through the built-in functions in Qiskit and Qutip \cite{Johansson2012}. Starting from the Choi matrix of the ideal and simulated gates, the diamond distance and the average gate fidelity can be evaluated through the Qiskit functions \texttt{average\_gate\_fidelity} and \texttt{diamond\_norm}. In particular, the latter makes use of the semidefinite program developed in Ref.~\cite{Watrous2012}. Finally, we have computed the unitarity of each gate by employing both definitions in Eqs.~\eqref{eqn:unitarity} and~\eqref{eqn:unitarityBis}. In particular, we have introduced two new functions in Qiskit to obtain these values starting from the Choi matrix of the process. We have checked that the values computed according to each definition coincide. 

Readout error mitigation has been performed for all the results shown in the paper. The calibration filters have been obtained through the class \texttt{CompleteMeasFitter} and the function \texttt{complete\_meas\_cal} in Qiskit.

\section{Further results}
\label{sec:further}

\subsection{Detrimental effects of noise}
\label{sec:detrimental}

\begin{figure*}
    \centering
  \includegraphics[scale=0.38]{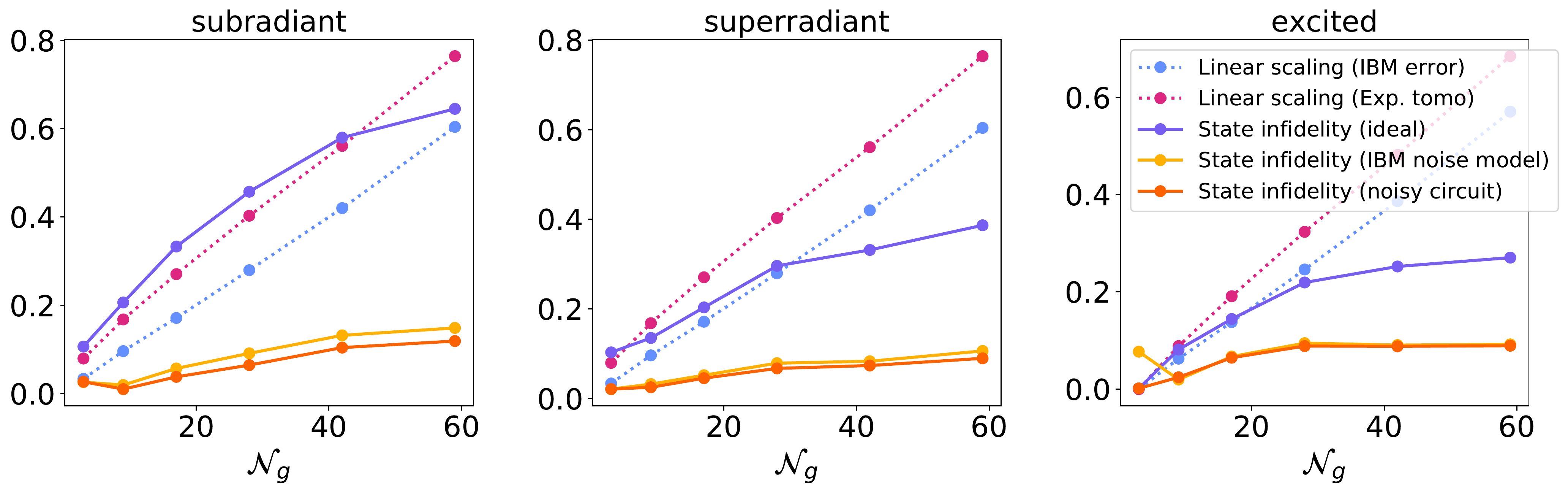}%
    \caption{Scaling of the error as a function of the number of CNOT gates in the algorithm $\mathcal{N}_g$ during the simulation of the collective dynamics displayed in Fig.~\ref{fig:countsTwo}. The markers indicate the steps of the collision model. Dotted light blue line and dotted magenta line: linear scaling of respectively the IBM gate error and the experimental average gate infidelity, given in Fig.~\ref{fig:erroran2}. Solid violet line: infidelity between the experimental state and the ideal one. The errors bars for the latter quantity have been obtained as the standard deviations of 100 realizations of a random sample of the state tomography data via bootstrapping, and they are within the markers. Solid orange line: infidelity between the experimental state and the state simulated on the noisy circuit. Solid yellow line: infidelity between the experimental state and the state simulated via the IBM noise model.}
    \label{fig:errortimestep2}
\end{figure*}

In this section, we will focus on the second set of results on the implementation of the MCM on \texttt{ibmq\_guadalupe}. We will observe how a pair of very noisy gates can jeopardize the simulation of collective effects on near-term devices.

The outcomes of the measurements in the {computational} basis at each step of the algorithm are shown in Fig.~\ref{fig:countsTwo}. It is quite evident that the results are very noisy, and almost no signature of collective effects can be extrapolated from them. In particular, the experimental evolution of the subradiant state following the collective dynamics (light blue markers)  shows no slower decay than in the case of local dissipation only (solid black lines). The superradiant state (magenta markers) does display a fast (although not always accurate) decay during the first two collisions, but decoherence gains the upper hand starting from the third one, and the dynamics stabilizes at a wrong value. The concavity of the evolution of $\ket{ee}$ (orange markers) is often erroneous as well.

The gate analysis in Fig.~\ref{fig:erroran2} may shed light on the origin of these noisy results. We have found a very large value of the experimental average gate infidelities of the CNOTs 1--2 and 2--1, which are repetitively employed in the algorithm, as explained in Appendix~\ref{sec:expScheme}. The corresponding IBM values are not as high. These gates may be responsible for the noisy dynamics in Fig.~\ref{fig:countsTwo}. Indeed, we can once again compare the experimental state with the states simulated through the ``noisy circuit'' noise model and the IBM noise model introduced in Sec.~\ref{sec:linearscaling}. This is done in Fig.~\ref{fig:errortimestep2}. We observe that, especially for the subradiant dynamics, the noise model based on the noisy circuit built through the experimental Choi matrices predicts a slightly better state than the IBM noise model. We have found that this is due to the fact that the IBM noise model is too optimistic, as it makes use of a relatively not-so-noisy CNOT gates 1--2 and 2--1. So, in the presence of some very noisy gates the full process tomography may actually be an improved tool to understand the errors in the quantum simulation.

It is worth stressing that the gate analysis for the second set of results confirms the noise properties we previously found for the first set. In particular, the source of noise is mostly incoherent, since the ratio between incoherence and gate infidelity is almost always larger than 0.9, as depicted in the upper plot of Fig.~\ref{fig:erroran2}. Moreover, the scaling of the state infidelity as a function of the channel length is again linear or sublinear. Finally, the diamond distance is of the same order of magnitude as for the first set of results, as discussed in Appendix~\ref{sec:diamond}.

\begin{figure*}
    \centering
  \includegraphics[scale=0.32]{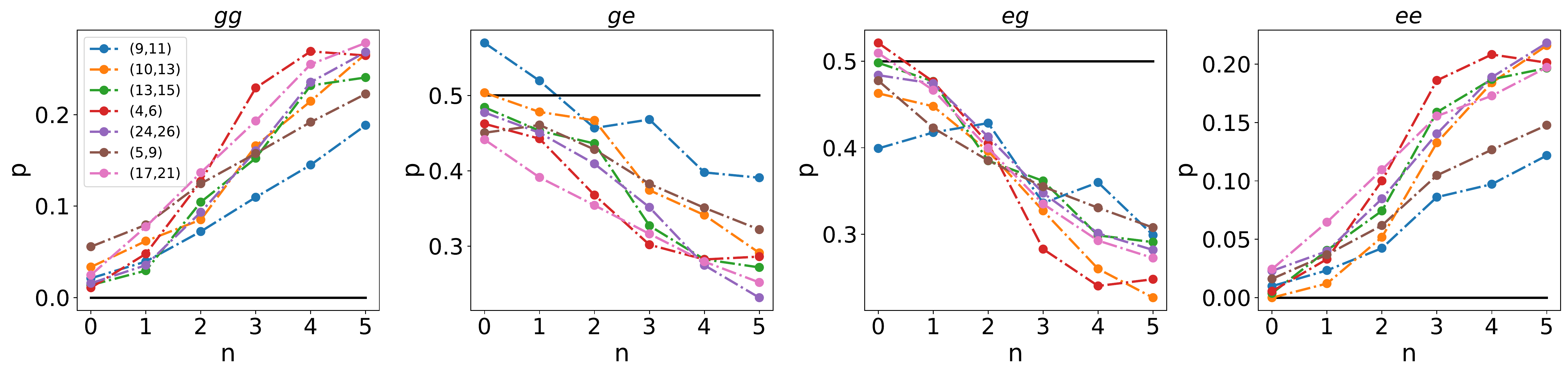}%
    \caption{Probability of finding the states $\ket{gg}$, $\ket{ge}$, $\ket{eg}$, $\ket{ee}$ (from left to right) through a projective measurement in the {computational} basis as a function of the number of steps in the MCM on \texttt{ibmq\_toronto}, when the initial state is the subradiant state. Solid black lines: theoretical prediction based on the master equation. Markers of the dash-dotted lines: experimental values. Different colors denote different initial pairs of system qubits in the topology of \texttt{ibmq\_toronto}, each of which interacts with a corresponding train of ancillas.}
    \label{fig:toronto}
\end{figure*}

\subsection{Subradiant dynamics on \texttt{ibmq\_toronto}}
\label{sec:toronto}
The validity of the experimental results presented in Sec.~\ref{sec:IBM} is not restricted to \texttt{ibmq\_guadalupe}, or to a specific choice of system and ancillary qubits. Indeed, in Fig.~\ref{fig:toronto} we show the experimental outcomes for the MCM implemented on \texttt{ibmq\_toronto}, which is a different IBM backend with 27 qubits. We have run the MCM to simulate the collective subradiant dynamics starting from different pairs of system qubits, each of which was connected to a suitable train of ancillas. The findings plotted in Fig.~\ref{fig:toronto} confirm the emergence of collective effects in the quantum simulation, as discussed in Sec.~\ref{sec:IBM} and Fig.~\ref{fig:countsOne}. Once again, the gate errors break the decoherence-free subspace of the ideal dynamics, but still, comparing Fig.~\ref{fig:toronto} with Fig.~\ref{fig:countsOne}, we observe a slower decay of the subradiant state than in the scenario with local dissipation only. In some cases, as for the system pair (9,11) (we refer the reader to the Qiskit documentation for the topology of the backends \cite{Qiskit}), the subradiance on \texttt{ibmq\_toronto} is actually enhanced with respect to the best results obtained on \texttt{ibmq\_guadalupe}. 

Due to the fact that \texttt{ibmq\_toronto} is re-calibrated much more often than \texttt{ibmq\_guadalupe}, it has not been possible to perform the process tomographies necessary for the gate analysis we introduced in Sec.~\ref{sec:IBM}. However, the crucial readout error mitigation has been correctly applied to the results in Fig.~\ref{fig:toronto}.

\subsection{Refreshing the ancilla through the reset gate}
\label{sec:reset}

\begin{figure*}
    \centering
  \includegraphics[scale=0.32]{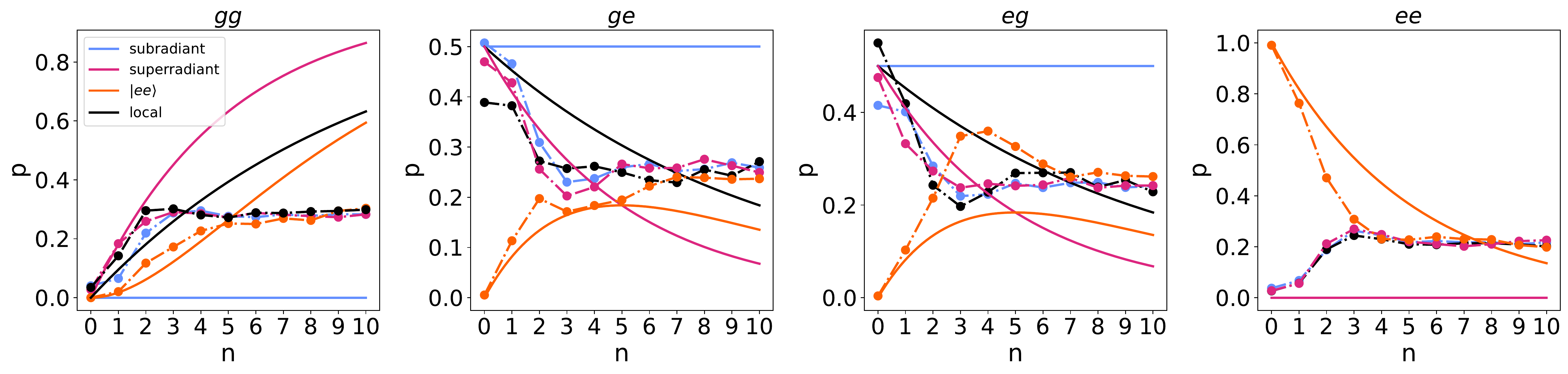}%
    \caption{Probability of finding the states $\ket{gg}$, $\ket{ge}$, $\ket{eg}$, $\ket{ee}$ (from left to right) through a projective measurement in the {computational} basis as a function of the number of steps in the MCM on \texttt{ibmq\_guadalupe}, when the initial state is the subradiant state (light blue), superradiant state (magenta), $\ket{ee}$ (orange), and when the dynamics is local starting from the subradiant state (black). Solid lines: theoretical prediction based on the master equation. Markers of the dash-dotted lines: experimental values. The reset gate is employed to refresh the state of the ancillary qubit after each timestep.}
    \label{fig:reset}
\end{figure*}

As discussed in Appendix~\ref{sec:expScheme}, a possible way to optimize the necessary resources  for the quantum simulation of the MCM is represented by the reset gate available on some IBM computers \cite{Qiskit}, which re-initializes the state of the target qubit in the ground state $\ket{0}$. With a reset gate at our disposal, we need a single ancillary qubit (two ancillary qubits in the case of the local decay) to simulate collective effects, since we can re-initialize it after each collision and avoid using a whole train of ancillas, which must be swapped. It goes without saying that this would represent a huge improvement in the number of necessary qubits and gates. Moreover, the ability to reset qubits in parallel with unitary gates would be beneficial for quantum error correction schemes, as it would provide a mechanism for flushing entropy out of system qubits.

We have run a protocol based on the MCM with the reset gate on \texttt{ibmq\_guadalupe}, and the results are shown in Fig.~\ref{fig:reset}. The possibility of using always the same ancilla allows us to implement more collisions (here, we have chosen $n=10$). However, the accuracy of the simulation is clearly much worse than for the MCM via trains of ancillas. Indeed, after one or two collisions, decoherence arises and the state of the system qubits reaches what looks like a thermal stationary state with no coherences at all. 

This is due to the fact that the running time of the reset gate is much longer than the one of standard single- and two-qubit gates \cite{Qiskit} (the reset gate length at the time of the experiment was around 7.34 $\mu$s, while the $T_1$ time of the ancillary qubit was around 80.0 $\mu$s, so 10 applications of the reset gate would already lead to complete dissipation). That is, implementing a dissipative channel on the quantum platform takes more time than performing coherent operations. Therefore, employing the reset gate even just a couple of times is enough for decoherence to occur. This being said, we point out that no dynamical decoupling scheme \cite{Viola1998} has been performed during the application of the long-duration reset gate. Dynamical decoupling may therefore improve the results in Fig.~\ref{fig:reset}.

In conclusion, at present (the results in Fig.~\ref{fig:reset} have been obtained during December 2021), the reset gate is not a feasible solution to reduce the number of necessary resources for the quantum simulation of the MCM. Similar results have been obtained on different IBM computers, including \texttt{ibmq\_toronto} and \texttt{ibmq\_mumbai}.

\section{Additional material on the noise analysis}
\label{sec:diamondTrace}

\begin{figure*}
\centering
  \includegraphics[width=.9\textwidth]{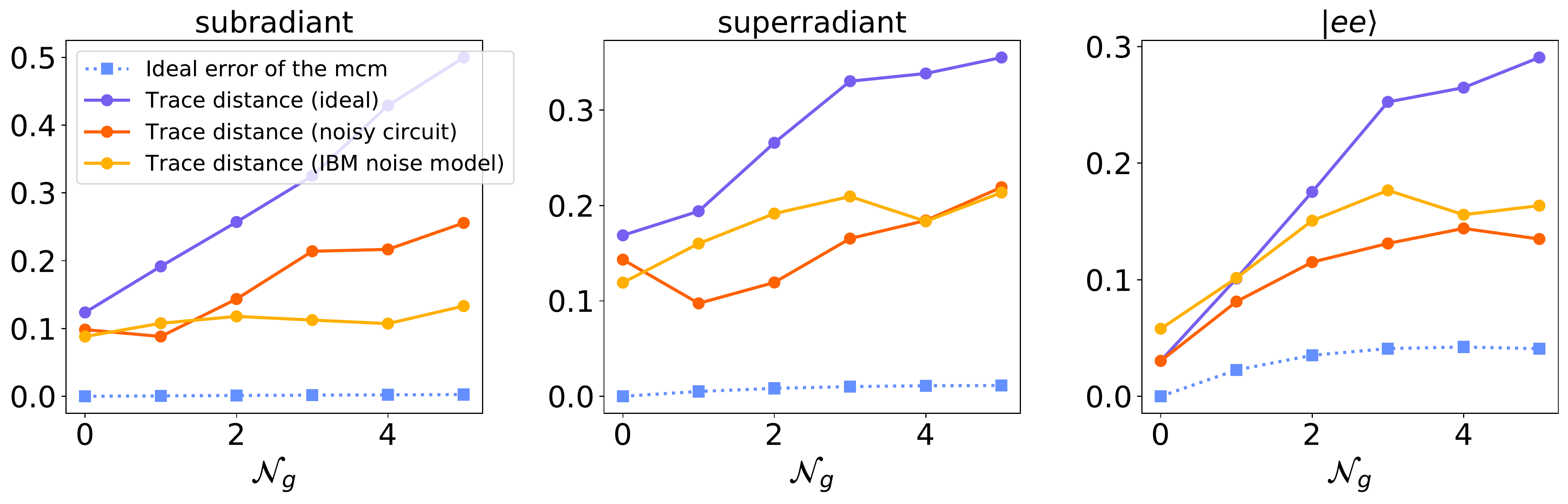}\\
  \includegraphics[width=.9\textwidth]{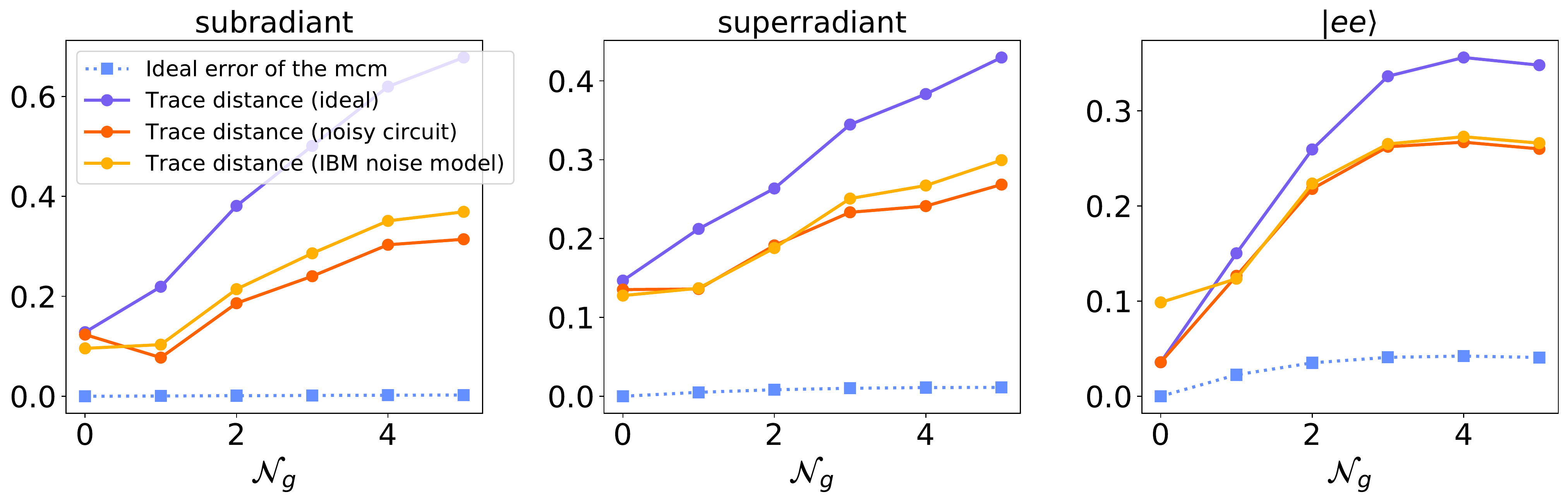}
    \caption{Trace distance between the experimental state and the ideal state of the algorithm (solid violet line), the simulated state through the noisy circuit (solid orange line) and the simulated state through the IBM noise model (solid yellow line), as a function of the number of gates in the algorithm $\mathcal{N}_g$, and for the first set of results in Fig.~\ref{fig:countsOne} (first row) and for the second set of results in Fig.~\ref{fig:countsTwo} (second row). The markers indicate the steps of the collision model. The dotted light blue line depicts the trace distance $\epsilon_{\text{ideal}}(n)$ between the state of the multipartite collision model at each timestep of the dynamics and the ideal physical state driven by the master equation, according to Eq.~\eqref{eqn:epsilonIdeal}.}
    \label{fig:traceDist}
   \end{figure*}
\subsection{Trace distance between experimental, ideal and simulated states}
\label{sec:trace}
Fig.~\ref{fig:traceDist} shows the trace distance between the experimental state and the ideal state of the algorithm and between the experimental state and the states simulated through the two noise models introduced in Sec.~\ref{sec:linearscaling}, as a function of the number of gates $\mathcal{N}_g$ and for each timestep of the algorithm (markers). The first row refers to the first set of results in Fig.~\ref{fig:countsOne}, while the second one to the second set of results in Fig.~\ref{fig:countsTwo}. We observe that for the first set of results the trace distances between experimental and simulated states have a similar behavior for the two different noise models; the IBM noise models outperforms the noisy circuit one for the subradiant dynamics, but it has a worse behavior for the superradiant and $\ket{ee}$ initial states. In contrast, the noisy circuit noise model is always better than the IBM noise model for the second set of results. Anyway, both noise models provide a good prediction for the experimental state of the dynamics. The trace distance has a larger value than the infidelity between the same states, depicted in Figs.~\ref{fig:errortimestep1} and~\ref{fig:errortimestep2}, because of the square root dependence between these quantites, expressed by Eq.~\eqref{eqn:fidelityBound}. 

The scaling of the trace distance between experimental and ideal states is linear or sublinear as a function of the number of gates. Finally, in Fig.~\ref{fig:traceDist} we also plot the quantity $\epsilon_{\text{ideal}}(n)$ given by Eq.~\eqref{eqn:epsilonIdeal}, that is, the trace distance between the ideal state of the multipartite collision model at the $n$th timestep and the ``physical'' state $\rho_S(n\Delta t)$ generated by the master equation~\eqref{eqn:LiouvillianSup}. As we may also deduce from the inset in Fig.~\ref{fig:ideal}(a), the ratio between the ideal error $\epsilon_{\text{ideal}}(n)$ and the ``experimental error'' due to noisy gates (solid violet line) is very small (around $10^{-2}$) apart from the case where the initial state is $\ket{ee}$. In this case, the ratio is around $10^{-1}$, so that the experimental errors are still playing a major role in the discrepancy between the physical quantum dynamics we aim to simulate ($\rho_S(n\Delta t)$) and its experimental implementation via MCM.  

\subsection{Diamond distance of the CNOT gates}
\label{sec:diamond}
\begin{figure*}
  \includegraphics[scale=0.3]{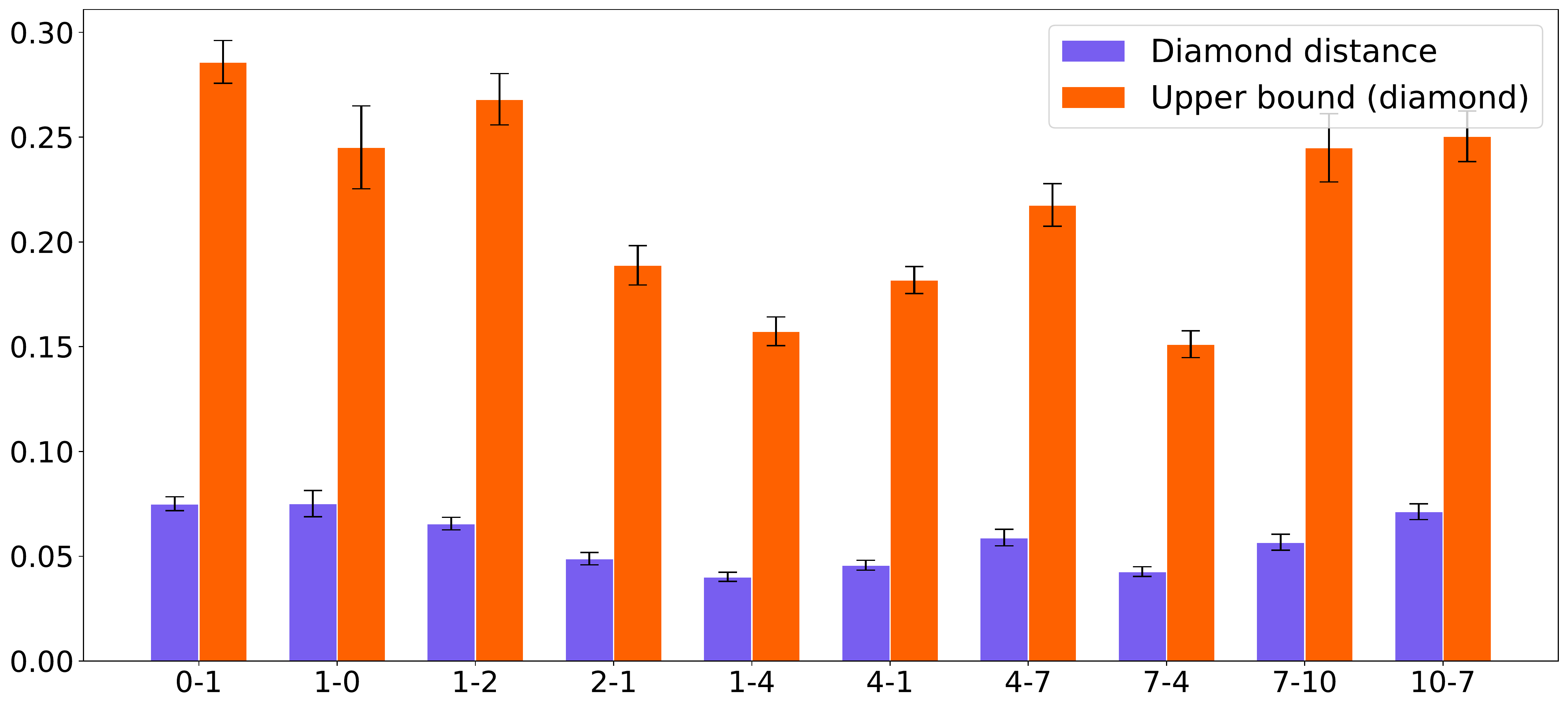}\\
  \includegraphics[scale=0.3]{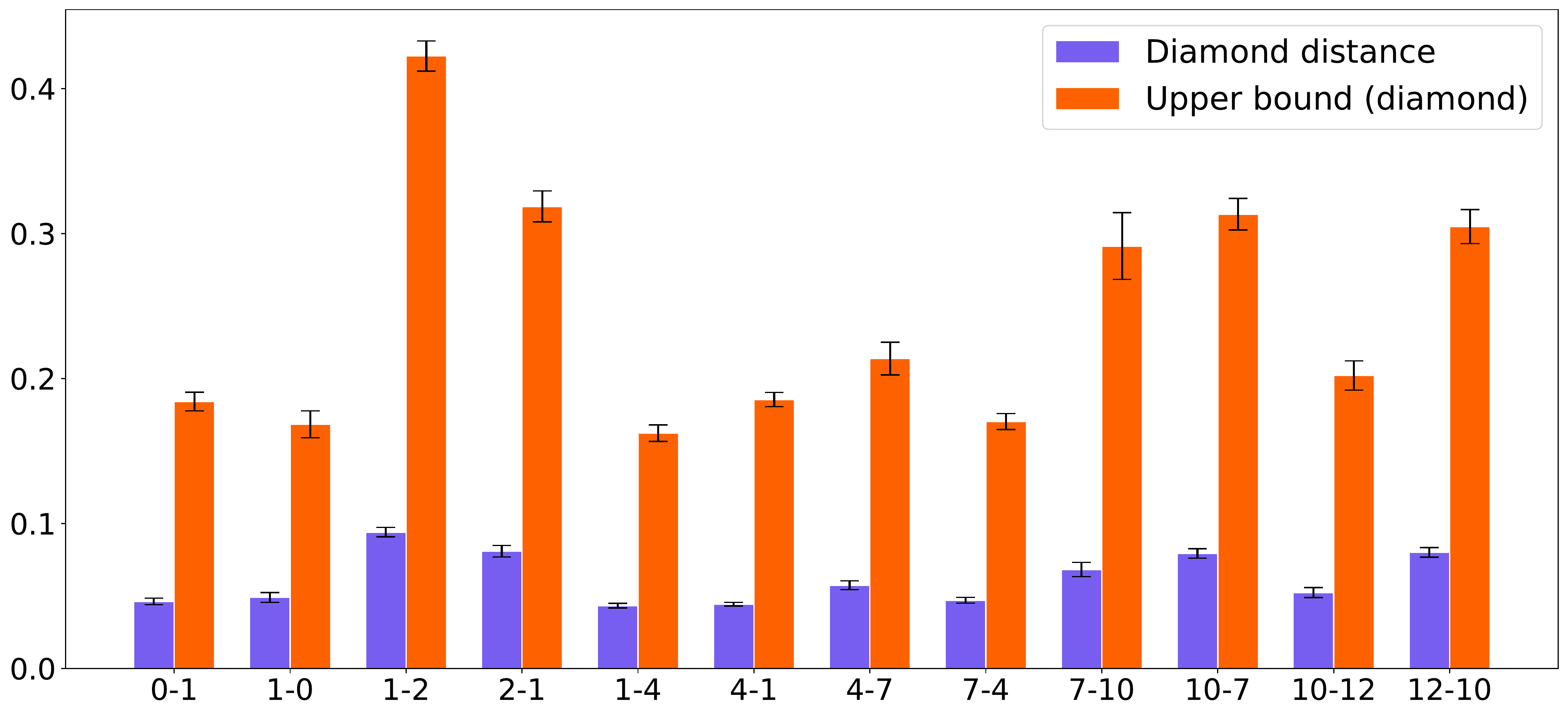}%
    \caption{Error analysis of the set of CNOT gates in \texttt{ibmq\_guadalupe} employed for the simulation of the first set of results in Fig.~\ref{fig:countsOne} (upper plot) and for the second set of results in Fig.~\ref{fig:countsTwo} (lower plot)}. The tick 0--1 on the x axis corresponds to the CNOT gate where ``qubit 0'' in \texttt{ibmq\_guadalupe} is the control qubit and ``qubit 1'' is the target. We plot the diamond distance between ideal and noisy CNOT according to Eq.~\eqref{eqn:diamondDistance} (violet), and the corresponding upper bound based on the average gate fidelity and unitarity (orange), as defined in Eq.~\eqref{eqn:upperBoundDiamondUnitarity}. The error bars  in the plots are given by the standard deviations of 100 realizations of a random sample of the experimental data via bootstrapping.
    \label{fig:erroranDiam1}
\end{figure*}

In Fig.~\ref{fig:erroranDiam1} we show the diamond distance between ideal and experimental gate for the CNOTs employed in the algorithm. The distances have been computed by reconstructing the Choi matrices of the experimental gates via process tomography, and the experimental values have been resampled 100 times via bootstrapping to estimate the standard deviations of the samples (error bars in Fig.~\ref{fig:erroranDiam1}). We observe that the experimental diamond distances for the noisy CNOTs are mostly between $5\times 10^{-2}$ and $10^{-1}$. As the threshold values for fault-tolerant quantum computation against general noise are tipically $\epsilon\approx 10^{-3}$--$10^{-4}$ \cite{Blume-Kohout2017a,Aliferis2009,Jones2018,Puri2020,Nielsen2021}, our experimental results seem to indicate that current near-terms quantum computers are still orders of magnitude away from these thresholds. Note, however, that the values in Fig.~\ref{fig:erroranDiam1} may still be biased due to residual SPAM errors \cite{Merkel2013}, which anyway have been mitigated as described in Sec.~\ref{sec:gateErrorAn}.

The values of the diamond distance in Fig.~\ref{fig:erroranDiam1} are  interesting for our purposes also because they are the diamond distances appearing in the bound in Proposition 2, Eq.~\eqref{eqn:boundNoisyMCM}, when we restrict ourselves to the errors on two-qubit gates only. This makes sense because, as stated before, the errors on single-qubit gates are typically around a couple of orders of magnitude smaller than on two-qubit gates. Moreover, the state preparation of the ancillas in the ground state at the beginning of the algorithm is very accurate on IBM quantum computers, whereas the swap gates to bring the new ancillas in contact with the system qubits are taken into account by the distances in the plot. Therefore, we can use the values in Fig.~\ref{fig:erroranDiam1} to estimate the upper bound on the simulation error of the MCM due to its noisy implementation via imperfect gates. As the diamond errors are quite large and we are using tens of CNOT gates in the algorithm, the bound in Eq.~\eqref{eqn:boundNoisyMCM} quickly exceeds 1.

For completeness, in Fig.~\ref{fig:erroranDiam1} we  plot also the upper bound on the diamond distance based on the infidelity and unitary, as given by Eq.~\eqref{eqn:upperBoundDiamondUnitarity}. We observe that the actual value of the diamond distance is always quite far from this upper bound (typically between three and five times smaller).

\bibliography{biblio}

\end{document}